\documentclass[3p]{elsarticle}

\usepackage{lineno}%,hyperref}
\usepackage{amsmath}
\usepackage{float}
\usepackage{graphicx}% Include figure files
\usepackage{dcolumn}% Align table columns on decimal point
\usepackage{bm}% bold math
\usepackage{amssymb}   % for math
\usepackage{tikz}
\usepackage{tikzorbital}
\usepackage{caption}
\usepackage{subcaption}
\usepackage{csquotes}
\usepackage{comment}% to use comment
\usepackage[outdir=./]{epstopdf}
\usepackage[multidot]{grffile}
\usepackage[colorlinks,linkcolor=green,urlcolor=green]{hyperref}
\DeclareUnicodeCharacter{2212}{-} 

\modulolinenumbers[5]

\biboptions{compress}
\begin{document}

\begin{frontmatter}

\title{Enhancement of maximum superconducting temperature by applying pressure and reducing the  charge transfer gap}

\author{Yi-Hsuan Liu$^{1,2*}$ , Huan-Kuang Wu$^{3}$ and Ting-Kuo Lee$^{2,4}$}
\address{$^1$ Department of Physics, National Tsing Hua University, Hsinchu
	30013, Taiwan}
\address{$^2$ Institute of Physics, Academia Sinica, Nankang 11529, Taiwan}

\address{$^3$ Department of Physics, University of Maryland, College Park, MD 20742, USA}

\address{$^4$ Department of Physics, National Sun Yat-sen University, Kaohsiung 80424, Taiwan}
\address{$^*$ yhliu0043@gapp.nthu.edu.tw}

\begin{abstract}
	Recent Scanning Tunneling Spectra(STS) measurement on underdoped cuprate discovers the increase of the  maximum superconducting transition  temperature $T_c$  when the size of charge transfer gap (CTG) is reduced. Applying pressure  is another well known method  to increase maximum $T_c$.  However, these pressure experiments also found another puzzle that $T_c$ is enhanced in underdoped and optimal doped samples but suppressed in overdoped. Here we present a possible mechanism based on the  charge fluctuation to explain both these two effects simultaneously. Starting from 3-band Hubbard model, we retrieve the charge fluctuation(CF) between oxygen 2$p^6$ band and copper 3$d^{10}$ band which is ignored in the $t-J$ model. This model is studied via variational Monte Carlo method(VMC). 
\end{abstract}


\end{frontmatter}

\section{Introduction}\label{introduction}
~~~~~~~About 30 years ago, right after  the discovery of high temperature superconductor\cite{Bednorz1986,PhysRevLett.58.908,RevModPhys.78.17,doi:10.1080/00018730701627707,Ogata2008} (HTS), Anderson\cite{ANDERSON1196} has proposed that this system is a Mott insulator at half filling. Either the strong coupling one-band Hubbard model or its equivalent low-energy $t-J$ model is a good starting Hamiltonian to understand the physics. Later, it was shown by Emery \cite{PhysRevLett.58.2794} that the half-filled  cuprate with both copper and  oxygens considered is actually a charge transfer insulator to be described by a three-band Hubbard model. Not long after, Zhang and Rice\cite{PhysRevB.37.3759,PhysRevB.41.7243} showed that the large charge transfer gap could help to form a  singlet state, known as Zhang-Rice singlet (ZRS),  representing the doped hole.  Therefore, the system could still be considered as an effective one-band model. In this case, the oxygen charge degrees of freedom are effectively frozen. Also, there are a number of experiments to support the presence of ZRS\cite{PhysRevB.88.134525,PhysRevLett.115.027002,PhysRevLett.95.177002}. %Interested reader are referred to the review articles\cite{}.

%(which is the t-J model).In this paper, we restrict the discussion in single-layer materials. ed2007

Many works tried to search for the dominant factors to enhance the superconducting transition temperature $T_c$, for example, the number of layers, the distance of apical oxygen, the magnetism, the hole density of Cu and O and the charge transfer gaps, etc. 
Few years ago, Weber et.al\cite{Weber2012,Weber2017102} have shown that the maximum $T_c $ or $T_c^{max}$ becomes larger with the  decreasing charge  transfer energy, increasing nearest hopping $|t'|$ and apical oxygen distance. 
Recently several experiments have provided evidences to reexamine this idea of one-band model beyond optimal doping\cite{Badoux2015,PhysRevB.95.224517,Doiron-Leyraud2017}. The charge transfer gap (CTG) was directly measured by using the scanning tunneling microscopy(STM) for one and two layers $Bi_2(Sr,La)_2CuO_{6+\delta}$(Bi-2201) and $Ca_2CuO_2Cl_2$(CCOC)\cite{Ruan2016}.  Not only the gap size is surprisingly small of the order of $1-2$ eV,  but the maximum superconducting temperature $T_c^{max}$ is higher for the smaller gap. The result is also confirmed by angle-resolved photoemission spectroscopy (ARPES)\cite{PhysRevB.96.245112}. Thus the magnitude of the CTG is  about same order of magnitude as the hopping energy between oxygen $p$ orbital and copper $d$ orbital, clearly we need to have a re-examination of the CF effect.

The effect of charge transfer between Cu3$d^{10}$ and O2$p^6$,  which shall be denoted as CF hopping in this paper, is also examined by a number of other experiments. For example, applying hydrostatic \cite{PhysRevB.52.6854,Chu1995} and uniaxial pressure\cite{PhysRevLett.105.167002}, will enhance $T_c^{max}$.  However, a puzzling phenomena \cite{Yamamoto2015} is the reduction of $T_c$  with pressure in the overdoped (OD) regime  but enhancement of $T_c$  in the underdoped (UD) and optimal doping regimes, it leads to the shift of the $T_c$ dome to lower dopant density and with enhancement of $T_c^{max}$. This unusual behavior for $T_c$ under pressure cannot be simply attributed to the change of lattice constant or carrier density.  A theoretical understanding is still lacking. We will show below that proper consideration of CF hopping provides a simple explanation of enhancement of $T_c^{max}$ and the oppostie behavior in the UD and OD regimes.
%\cite{Weber2012,Weber2017102}.

Before we start to present  the details on our model and calculations, we note that 
the model without explicitly including oxygen has successfully explained a number of experiments. However, to study the CF effect, we will have to bring back the oxygen degrees of freedom. Instead of taking into account of the full degree of freedom of oxygen which will require extensive numerical works, in this paper we will present an effective model that keeps the ZRS but includes the CF effect to reveal the physics.

We will start with the Emery's 3-band mode~\cite{PhysRevLett.58.2794}:\\

\begin{equation}\label{eq:3bmd1}
\hat H_U = \sum_{l,\sigma} \epsilon_p p^{\dagger}_{l,\sigma}p_{l,\sigma}
+\sum_{i,\sigma} \epsilon_d d^{\dagger}_{i,\sigma} d_{i,\sigma}
+\sum_{i,\sigma} U_d \hat n^d_{i\uparrow} \hat n^d_{i\downarrow}
\end{equation}
\begin{equation}\label{eq:3bmd2}
\hat H_{pd} = -\sum_{i,\sigma}\sum_{l\subset i} t_{pd}V_{il} d^{\dagger}_{i,\sigma} p_{l,\sigma} + h.c
\end{equation}
where $l$ runs over the four oxygen around a copper, and $V_{il} = 1$ for $l = i+\frac{1}{2}x$ and $i+\frac{1}{2}y$, while $V_{il} = -1$ for $l = i-\frac{1}{2}x$ and $i-\frac{1}{2}y$. $d_{i,\sigma}^{\dagger}$ creates a d-hole with spin $\sigma$ in the Cu($d_{x^2 - y^2}$) at site i, $p_{j,\sigma}^{\dagger}$ creates a p-hole with spin $\sigma$ in the O($p_x$ and $p_y$) orbitals at site j. 
$\epsilon_d$ and $\epsilon_p$ correspond to the energy of the local Cu($d_{x^2 - y^2}$) and O($p_x$ and $p_y$) orbital, respectively.
$U_d$ is on-site Coulomb repulsion, of copper. The vacuum is defined as Cu~$d^{10}$ and O~$p^{6}$. 
%Here we use electron picture instead of hole picture since it is easier to understand how the electron fluctuate between oxygen and copper.

At half-filling, i.e., before a hole or an electron is doped into the system, every Cu will be at $d^{9}$ state with a single hole and all oxygens are at $p^{6}$ without holes and the system is a Mott insulator and a spin-$1/2$ quantum Heisenberg system \cite{ANDERSON1196}.
To consider the effect of doping, Zhang and Rice\cite{PhysRevB.37.3759,PhysRevB.41.7243} have derived an effective single-band model from Emery's 3-band model. In their argument , when a hole is doped into a cuprate, it will reside on oxygen. However, the superexchange coupling between this hole's spin and the spin at the nearest neighbor Cu~$d^{9}$ site will form a ZRS spin singlet.   All the other triplet and nonbonding state  have much higher energies and could be neglected from the low energy Hamiltonian.  In this picture, the hole's energy on oxygen site is actually the ZRS state with energy $\tilde \epsilon^{~}_p$, where the tilde means a renormalized energy of $\epsilon_p$. If the only charge degrees of freedom considered is from the doped holes that forming ZRS, then the degrees of freedom left are  the original spin -$1/2$ on the Cu $d^{9}$ site  and doped holes or ZRSs and then we obtain the well-known $t-J$  Hamiltonian\cite {PhysRevB.41.7243}.

\begin{equation}
\hat H_{t-J} = -t\sum_{\langle i, j\rangle} P_G c^{\dagger}_i c_j P_G + J\sum_{\langle i, j\rangle}(\vec{S_i}\cdot \vec{S_j} - \frac{1}{4}n_in_j)
\end{equation}
where $P_G = \prod_{i} (1 - \hat n_{i\uparrow}\hat n_{i\downarrow})$, $t$ is the effective hopping amplitude of ZRS, and $J$ is the superexchnage interaction between nearest neighbor Cu $d^{9}$ holes.

\begin{figure*}
	\centering
	\begin{subfigure}{0.70\linewidth}
		\centering
		\includegraphics[scale=0.5]{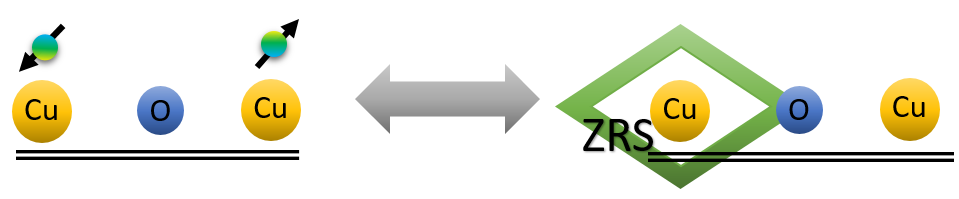}
		\caption{Three band picture  of CF hopping.  }
		\label{sfig:tbarhop}
	\end{subfigure}\par%\hfill
	
	\begin{subfigure}{0.70\linewidth}
		\centering
		\includegraphics[scale=0.5]{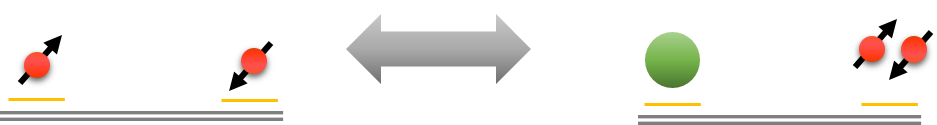}
		\caption{CF hopping in one band representation.  }
		\label{sfig:tbarhop1band}
	\end{subfigure}\par%\hfill
	
	\begin{subfigure}{0.70\linewidth}
		\centering
		\includegraphics[scale=0.5]{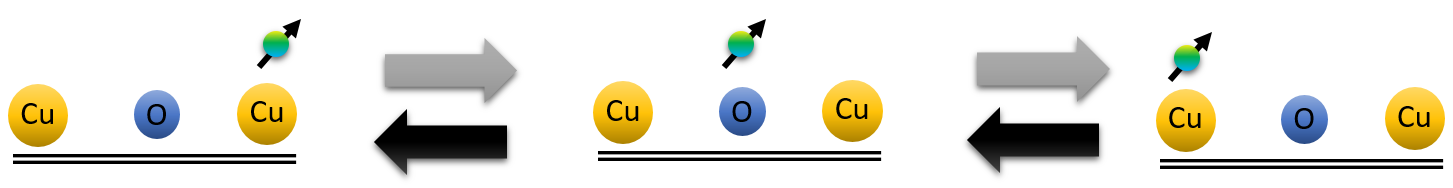}
		\caption{ Three band picture of doublon hopping.}
		\label{sfig:tdhop}
	\end{subfigure}%\hfill
	
	\begin{subfigure}{0.70\linewidth}
		\centering
		\includegraphics[scale=0.5]{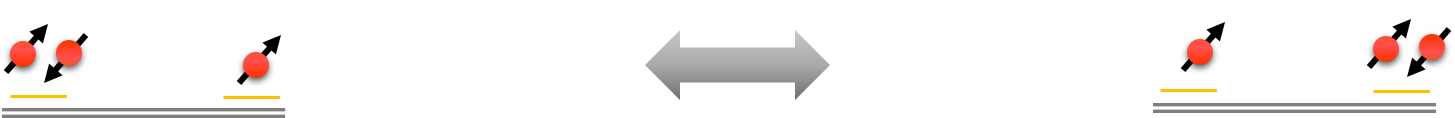}
		\caption{ Doublon hopping in one band representation.}
		\label{sfig:tdhop1band}
	\end{subfigure}%\hfill
	\caption{ Mapping the CF hopping processes in three bands into our effective one-band model with schematic illustrations. (a)The initial state has an O$2p^6$ in between two neighboring Cu$3d^9$ holes with opposite spin. In the three-band model, one of these two holes on Cu site could hop to the middle O site to form a ZRS with the hole on the other Cu and leaving a Cu $3d^{10}$ or the doublon state.  Hence this is equivalent to the formation of a holon-doublon pair in our one-band representation as shown in (b). (c) In the three-band model, the initial configuration of Cu$3d^{10}$, O$2p^6$ and Cu$3d^9$ could change to the intermediate configuration with Cu$3d^{10}$, O$2p^5$ and Cu$3d^{10}$, before it reaches the final configuration with Cu$3d^9$ O$2p^6$ and Cu$3d^{10}$, but this process is equivalent to the exchange between Cu $3d^{10}$ and neighboring Cu $3d^9$ which could be easily seen as the hopping of doublon in our representation shown in (d). }
	\label{fig:hoppingPic}
%	\raggedright
\end{figure*}

However, in principle several other charge degrees of freedom could be important. One possibility is that instead of being at the oxygen, the hole could sit at the Cu site to form Cu $d^{8}$ which has energy $2\epsilon_d +U_d$ according to Eq. (1).  This could be neglected at large $U_d$.  The other possibility is  to have the hole hopping from the Cu $d^{9}$ site to the nearest oxygen site  to form a ZRS with the adjacent Cu while leaving a Cu $d^{10}$ behind.  This is schematically shown in Fig.\ref{sfig:tbarhop}.  The energy difference between the initial and final configurations of this process is $\tilde \epsilon^{~}_p$ - $\epsilon_d$, which is exactly  the charge transfer gap, $\Delta_{CT}$, discussed before. If only this process is to be considered besides $t$ and $J$, we could completely eliminate the   oxygen and denote the two configurations as in Fig.\ref{sfig:tbarhop1band}, where the Cu $d^{10}$ is now represented by a doublon. The rate of this process switching from a pair of nearest-neighbor holes to a holon-doublon pair is denoted as  $\bar t$. Once we have Cu $d^{10}$ as a doublon, then it can also exchange with a Cu $d^{9}$ through the middle oxygen as shown in Fig.~\ref{sfig:tdhop}. This is equivalent to the hopping of a doublon, as shown in  Fig.\ref{sfig:tdhop1band}.
Finally, we have an effective one-band charge transfer model
\begin{align}\label{eq:cfmodel}
\hat H_{CT} = \hat H_{t-J} + \hat H_{CF}
\end{align}
\begin{equation} \label{eq:HCF}
\begin{split}
\hat H_{CF} =& -\bar t \sum_{\langle i, j\rangle\sigma} 
(1 - \hat n_{i,\bar \sigma})c_{i\sigma}^{\dagger}c_{j,\sigma} \hat n_{j,\bar \sigma} + \hat n_{i,\bar \sigma}c_{i\sigma}^{\dagger}c_{j,\sigma}(1 - \hat n_{j,\bar \sigma}) + h.c\\
& -t_d\sum_{\langle i, j\rangle\sigma} \hat n_{i,\bar \sigma}c_{i\sigma}^{\dagger}c_{j,\sigma} \hat n_{j,\bar \sigma}  + h.c\\
& + \Delta_{CT}\sum_{i}\hat n_{i,\uparrow}\hat n_{i,\downarrow}
\end{split}
\end{equation}

where $\langle i,j\rangle$ denotes nearest neighboring sites. 
There are three kinds of nearest neighboring hoppings as shown schematically in Table.\ref{tab:hop}. First, $t$ is the hopping of ZRS or hole  in the $t-J$ model. Second, $\bar t$  controls the creation and annihilation of a holon-doublon pair. Finally, $t_d$ describes  doublon hopping.  We notice that the process of creation and annihilation of a holon-doublon pair is actually one of the many intermediates states considered in the derivation of the superexchange $J$\cite{ANDERSON1196, PhysRevB.37.3759,PhysRevB.41.7243}. Thus in principle $J$ should be reduced from the values used in the pure $t-J$ model by a small amount. Thus in this work we consider a small $J/t=0.33$. The generic parameter $J$ ranges from $0.1~eV$ to $0.13~eV$ among various cuprates, which has been determined systematically by Raman scattering\cite{PhysRevLett.60.732,PhysRevB.53.R11930} and neutron scattering\cite{PhysRevLett.79.4906, PhysRevLett.67.3622}.

%In this work, we also considered hopping $t'$ between second nearest neighbors. 
 
By setting $t=\bar t = t_d$, the model is equivalent to the $t-J-U$ model where $\Delta_{CT}$ is the effective Hubbard $U$.  The model\cite{PhysRevLett.84.4188, PhysRevB.63.100506, Laughlin2002,PhysRevLett.90.207002,Coleman2003,PhysRevB.79.014524} has been studied via various method like variational Monte Carlo (VMC) method \cite{PhysRevLett.90.187004,PhysRevB.72.045130,PhysRevB.79.144526,PhysRevB.90.115137}, slave-particle method\cite{1605.03969,0953-8984-28-11-116001} , Gutzwiller renormalized mean field theory(RMFT) \cite{PhysRevB.71.014508,PhysRevB.71.104505,PhysRevB.88.094502,Liu2014123,0253-6102-57-4-29,0953-8984-23-49-495602,Abram_2017}, diagrammatic expansion of the Gutzwiller wave function (DE-GWF)\cite{PhysRevB.96.054511,PhysRevB.95.024507,PhysRevB.95.024506} , and density matrix renormalization group(DMRG)\cite{PhysRevLett.84.4188,PhysRevB.96.205120}.
To examine CF effect more carefully, we will consider more general cases where the three hopping amplitudes, $t$, $\bar t$ and $t_d$, are different.  Their  difference will become apparent when we discuss the effect of pressure in Sec.\ref{sec:xxx}. The derivation of these parameters from the 3-band model of Eqs. (1) and (2) are quite tedious\cite{YHL2016}. They depend on CTG, hybridization $t_{pd}$ and $d-d$ Coulomb repulsion $U_d$.  In this work, we will just treat these as parameters of our model.

%We should note that t-J-U model has been studied via the various method like VMC\cite{PhysRevLett.90.187004,PhysRevB.72.045130,PhysRevB.79.144526,PhysRevB.90.115137}, slave-particle method\cite{1605.03969,0953-8984-28-11-116001} , Gutzwiller renormalized mean field theory(RMFT) \cite{PhysRevB.71.014508,PhysRevB.71.104505,PhysRevB.88.094502,Liu2014123,0253-6102-57-4-29,0953-8984-23-49-495602,Abram_2017}, diagrammatic expansion of the Gutzwiller wave function (DE-GWF)\cite{PhysRevB.96.054511,PhysRevB.95.024507,PhysRevB.95.024506} , and density matrix renormalization group(DMRG)\cite{PhysRevLett.84.4188,PhysRevB.96.205120}.
%Recently, a method of implementing RVB mean field and dynamical mean field theory(DMFT) is implemented on Bethe lattice\cite{PhysRevLett.117.136601}.
 
 %We also notice that E.Plekhanov et al.\cite{PhysRevLett.90.187004}  and S.Daul  \cite{PhysRevLett.84.4188} has pointed out that pair-pair correlation can be enhanced by additional exchange interaction $4t^2/U$ in the t-J-U model in UD and optimal dopant. However, the fundamental parameter that control pairing is unexplored previously. Furthermore, physics for charge transfer is not only described by CTG but also $\bar t$ and $t_d$.  

\begin{table}[h]
	\begin{center}
		\begin{tabular}{l|l|l|l}
			\hline
			t & $\underline{\uparrow}$~$\underline{0}$ & $\Longleftrightarrow$ & 0~$\underline{\uparrow}$ \\
			$\bar t$ & $\underline{\uparrow}$~$\underline{\downarrow}$& $\Longleftrightarrow$ & $\underline{0}$~$\underline{\uparrow\downarrow}$\\
			$t_d$& $\underline{\uparrow\downarrow}$~$\underline{\downarrow}$& $\Longleftrightarrow$ & $\underline{\downarrow}$~$\underline{\uparrow\downarrow}$
		\end{tabular}
	\end{center}
	\caption{Classification of hopping}\label{tab:hop}
\end{table}

 This paper is organized as follows. In Section.\ref{sec:Formalism} the wave function used in the VMC method  is presented. The effect of the three CF parameters: $\bar t$, $t_d$ and $\Delta_{CT}$  are analyzed in Section.\ref{ssec:1}. 
 In Section.~\ref{sec:ctgtc} we discuss the relation between CTG and pairing magnitude. In Section \ref{sec:xxx}, we discuss the  effect of pressure on pairing. 
 Finally, the conclusion is given in Section. \ref{conclusion}.

\section{Formalism and Method}\label{sec:Formalism}

~~~~~~~For the variational ground state ansatz in the VMC calculation, we use the correlated $d$-wave BCS wave function \cite{refId0,doi:10.1143/JPSJ.65.3615,PhysRevB.43.12943,Shih2004,PhysRevB.70.220502}
\begin{equation}
|\Psi\rangle = \hat P_G \hat P_{dh} \hat P_{N_e}|\Psi_{d-BCS}\rangle
\end{equation}
where $\hat P_{Ne}$ is the projection operator to fix the total number of particles to be $N_e$ and  the $d$-wave BCS trial wave function\cite{refId0,doi:10.1143/JPSJ.65.3615} is 
\begin{equation}
\mid \Psi_{d-BCS} \rangle =  \prod_{k\subset BZ}(u_k + v_k c^{\dagger}_{k\uparrow}c^{\dagger}_{-k\downarrow})|0\rangle
\end{equation}
where $u_k = \frac{1}{2}({1 + \frac{\xi_k}{\sqrt{\xi_k^2 + \Delta_k^2}}})$, $u_k^2 + v_k^2 = 1$,$\xi_k = -2(cosk_x+ cosk_y) - 4t_v'cosk_xcosk_y-\mu_v$ , 
$\Delta_k = \Delta_v(cosk_x - cosk_y)$, $t_v^{\prime}$, $\mu_v$ and $\Delta_v$ are  variaitonal parameters.
$\hat P_G$ is the well-known Gutzwiller projector \cite{PhysRevLett.10.159}
\begin{equation}
\hat P_G = \prod_i(1 - (1-g)\hat n_{i\uparrow}n_{i\downarrow})
\end{equation}
which controls the doublon density. For $ g = 0 $, there is no doublon and we have the strongest correlation in the large $\Delta_{CT}$ limit and this is used for the pure $t-J$ model. On the contrary, there is  no correlation for $g = 1$.	
 The Jastrow-type factor \cite{PhysRev.98.1479} $\hat P_{dh}$ is introduced to provide an attractive interaction between doublon and hole\cite{doi:10.7566/JPSCP.3.015012,doi:10.7566/JPSJ.82.014707,PhysRevB.95.035133} to make sure at half-filling we recover the Mott  insulator without free carrier, it is of the form 
 %Also, it was pointed out that the formation of bound exciton of spin singlet, or doublon-hole pair, is the elementary excitation inside the charge transfer insulator\cite{PhysRevB.58.13520}.
\begin{equation}
\hat P_{dh} = \prod_{i}(1 - Q_{dh}\hat B_i)
\end{equation}
where $\hat B_i = \prod_{\tau}\hat d_i(1 - \hat h_{i+\tau}),\tau = \hat x, \hat y$  represents a nearest neighbor holon-doublon binding state, $\hat d_i$ is the doublon creation operator $\hat d_i = c^{\dagger}_{i\uparrow}c_{i\uparrow}c^{\dagger}_{i\downarrow}c_{i\downarrow}$ and
$\hat h_i$ is the hole creation operator $\hat h_i = (1-c^{\dagger}_{i\uparrow}c_{i\uparrow}) (1-c^{\dagger}_{i\downarrow}c_{i\downarrow})$. For  $Q_{dh} = 0$, doublon and hole are free, while for $Q_{dh} =1$ doublon cannot be separated from a hole. For $Q_{dh} < 0 $ doublon and hole are repulsive to each other. In our case, for large $\Delta_{CT}$, $Q_{dh}$ is always positive.

%	\begin{equation}\end{equation}	

In order to find the ground state we optimize the trial wavefunction by implementing the stochastic reconfiguration(SR) method described in detail in Ref.\cite{PhysRevB.71.241103}

The superconductivity is characterized by the d-wave pairing order parameter
\begin{align*}
\Delta_{SC} = \sum_{i}<c_{i\uparrow}c_{i+x\downarrow}> - (x\leftrightarrow y)
\end{align*}
However, in canonical VMC, what we can measure is the pair-pair correlation function
\begin{align*}
P(\vec{r}) = \frac{1}{N_s}\sum_i\Delta^{\dagger}(\vec{R_i})\Delta(\vec{R_i}+\vec{r})
\end{align*}
where $N_s$ is the number of the sites and
\begin{align*}
\Delta(\vec{R_i}) = \frac{1}{\sqrt{2}}(\langle c_{i\uparrow}c_{i+x\downarrow} - c_{i\downarrow}c_{i+x\uparrow} \rangle - (x\leftrightarrow y))
\end{align*}
By taking the square root of the pair-pair correlation at the maximum distance, $r_{max}$, we can estimate the magnitude of pairing order parameter.
\begin{align*}
\Delta_{SC} \simeq \sqrt{P(\vec{r}_{max})}
\end{align*}

%Also we note that to estimate the error of $\Delta_{SC}$, we should take square root of mean value and the square of mean value when evaluate the variance.
%\clearpage

\section{Results}\label{result}	
~~~~~~~As mentioned above, the three hopping amplitudes in Table.\ref{tab:hop} are not necessarily the same. In this section, we investigate the effect of varying $\bar t$, $t_d$ and $\Delta_{CT}$ with fixed $t = 1$ and $J = 0.33$. Throughout this work, the lattice size considered is $20\times20$.
In the main text we will  show the result of having only the nearest neighboring hopping terms. A very similar result is obtained by  including the second nearest neighboring hopping terms where we set $t'/t=\bar t'/\bar t=t_d'/t_d=-0.3$ which is presented in the Supplementary Material. 

\subsection{Tuning parameters $\bar t$, $t_d$ and $\Delta_{CT}$}\label{ssec:1}

\begin{figure}
	\centering
	\begin{subfigure}{0.45\textwidth}
		\centering
		\begin{tikzpicture}
		\node[inner sep=1] (image) {\includegraphics[width=\linewidth]
		{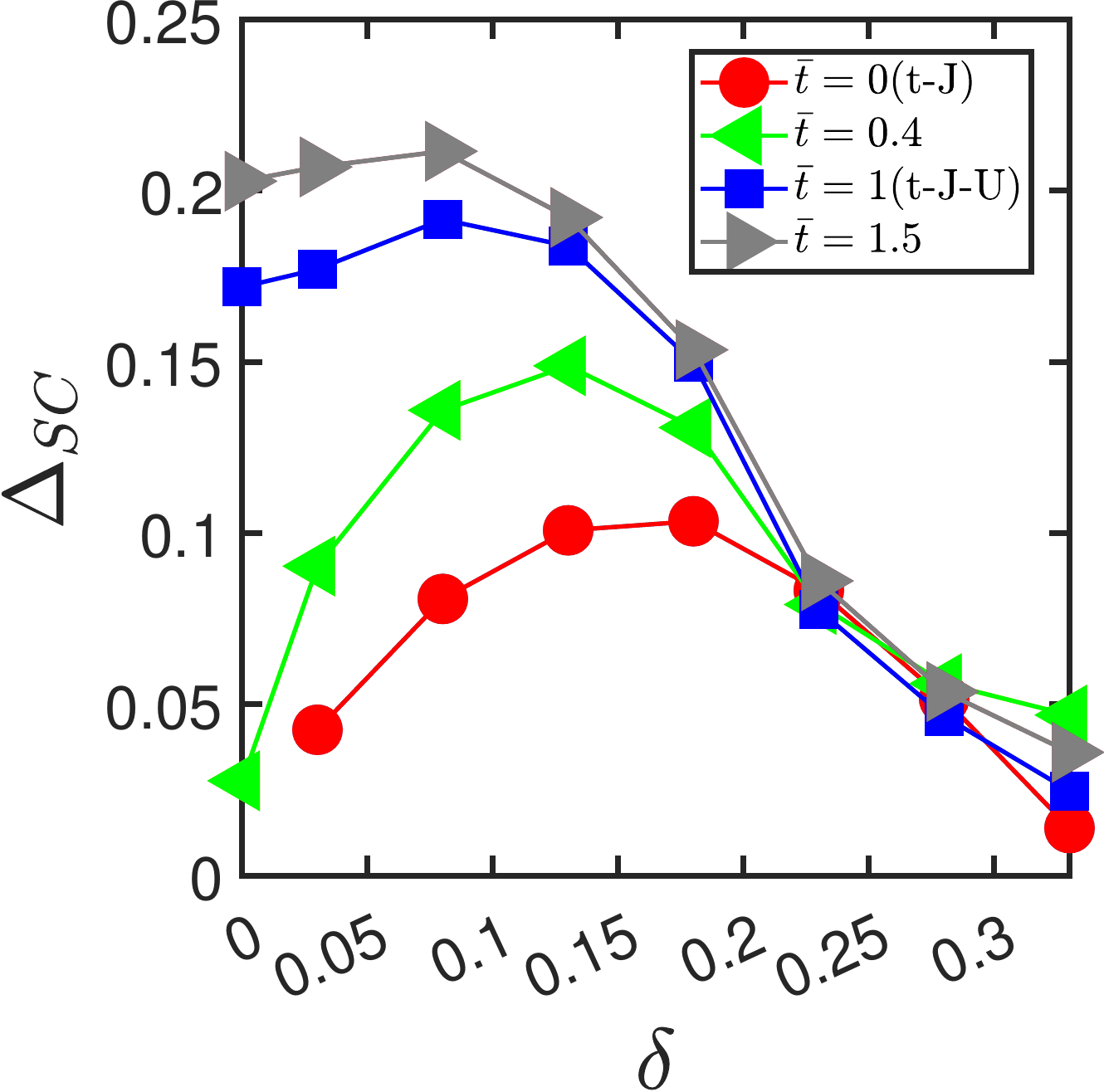}}; 
	%{./Resulteffecttune/2Ub6tp0OP}};
		\end{tikzpicture}
		\caption{}
		\label{sfig:efftbar1}	
	\end{subfigure}
	\hspace{1em}
	\begin{subfigure}{0.45\textwidth}
		\centering
		\begin{tikzpicture}
		\node[anchor=south west,inner sep=1] (image) {\includegraphics[width=\linewidth]
		{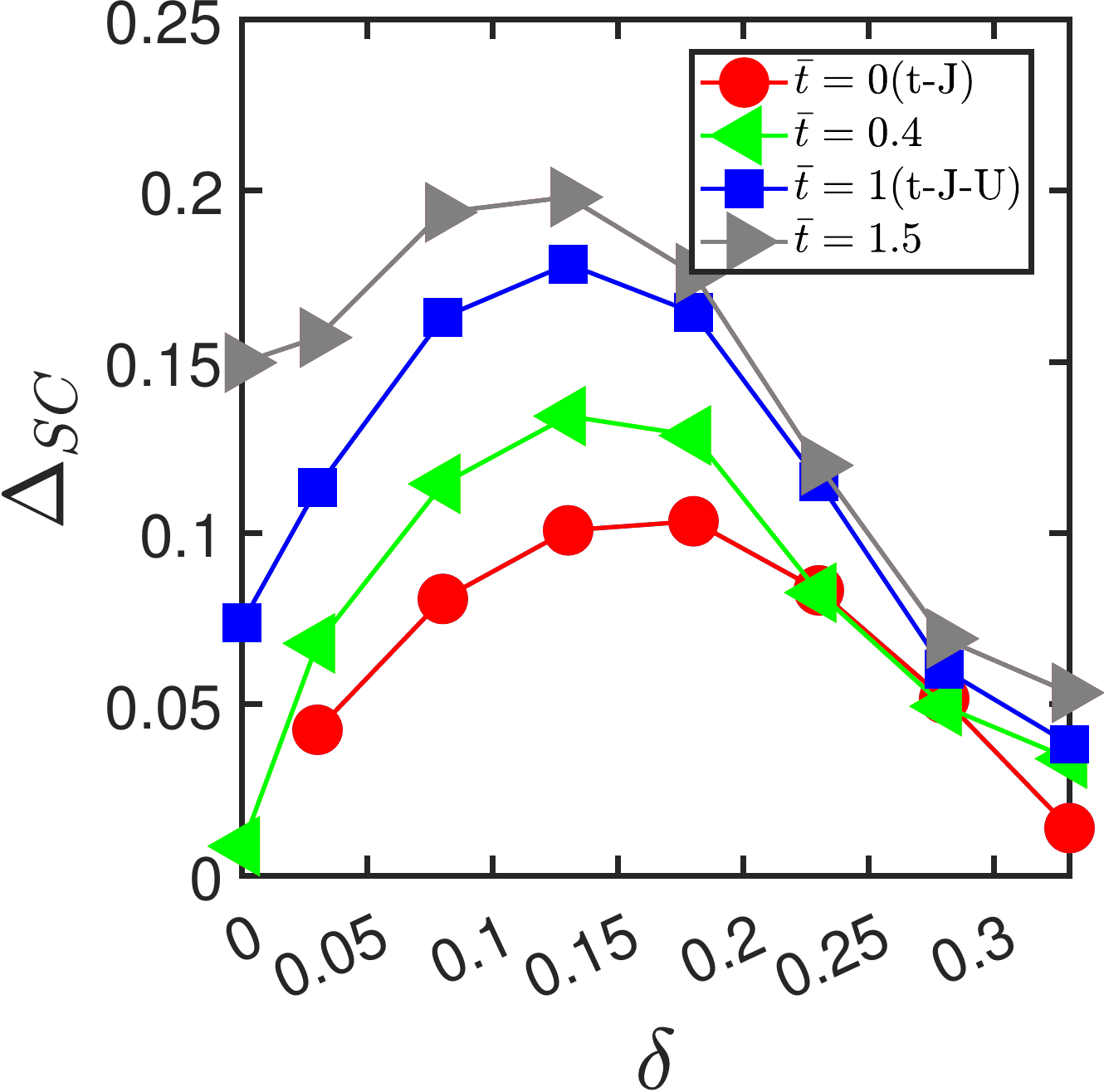}}; 
		%{./Resulteffecttune/2Ub8tp0OP}};
		\end{tikzpicture}
		\caption{}
		\label{sfig:efftbar2}	
	\end{subfigure}
	\caption{Pairing order parameter as a function of dopant concentration for various $\bar t$. Here we have $t_d=1$. The CTG is set to (a)$\Delta_{CT} = 6$ and  (b)$\Delta_{CT} = 8$.  Note that the label $\bar t = 0$ stands for $\bar t = t_d = 0$ which is the same as the $t-J$ model. For the case $\bar t=1$, we have the $t-J-U$ model.
	}%, and all the second nearest neighbor hopping terms $t'=t_d'=\bar t'=0$	
	\label{fig:efftbar}	
\end{figure}

\begin{figure}
	\centering
	\begin{subfigure}[b]{0.45\textwidth}
		\begin{tikzpicture}
		\node[inner sep=1] (image) {\includegraphics[width=\linewidth] {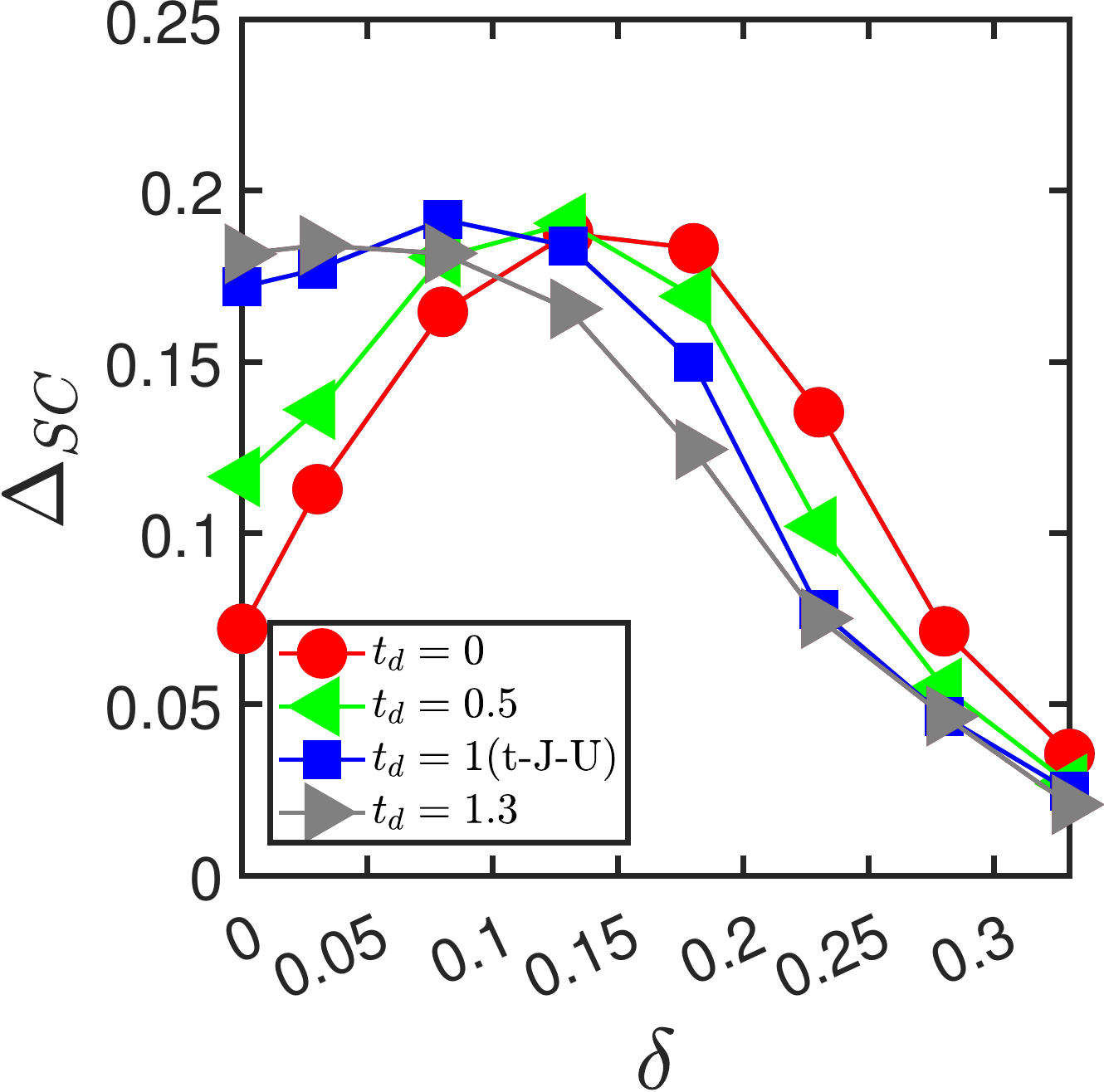}};;
		%{./Resulteffecttune/1Ub6tb1tp0OP}};;
		\end{tikzpicture}
		\caption{}
		\label{sfig:efftd1}	
	\end{subfigure}
	~
	\begin{subfigure}[b]{0.45\textwidth}
		\begin{tikzpicture} 
		%\node[anchor=south west,inner sep=0] (image) {\includegraphics[width=\linewidth] {./paper_D1/0RMFT_ppcd}};
		\node[anchor=south west,inner sep=1] (image)
		 {\includegraphics[width=\linewidth]
		 	{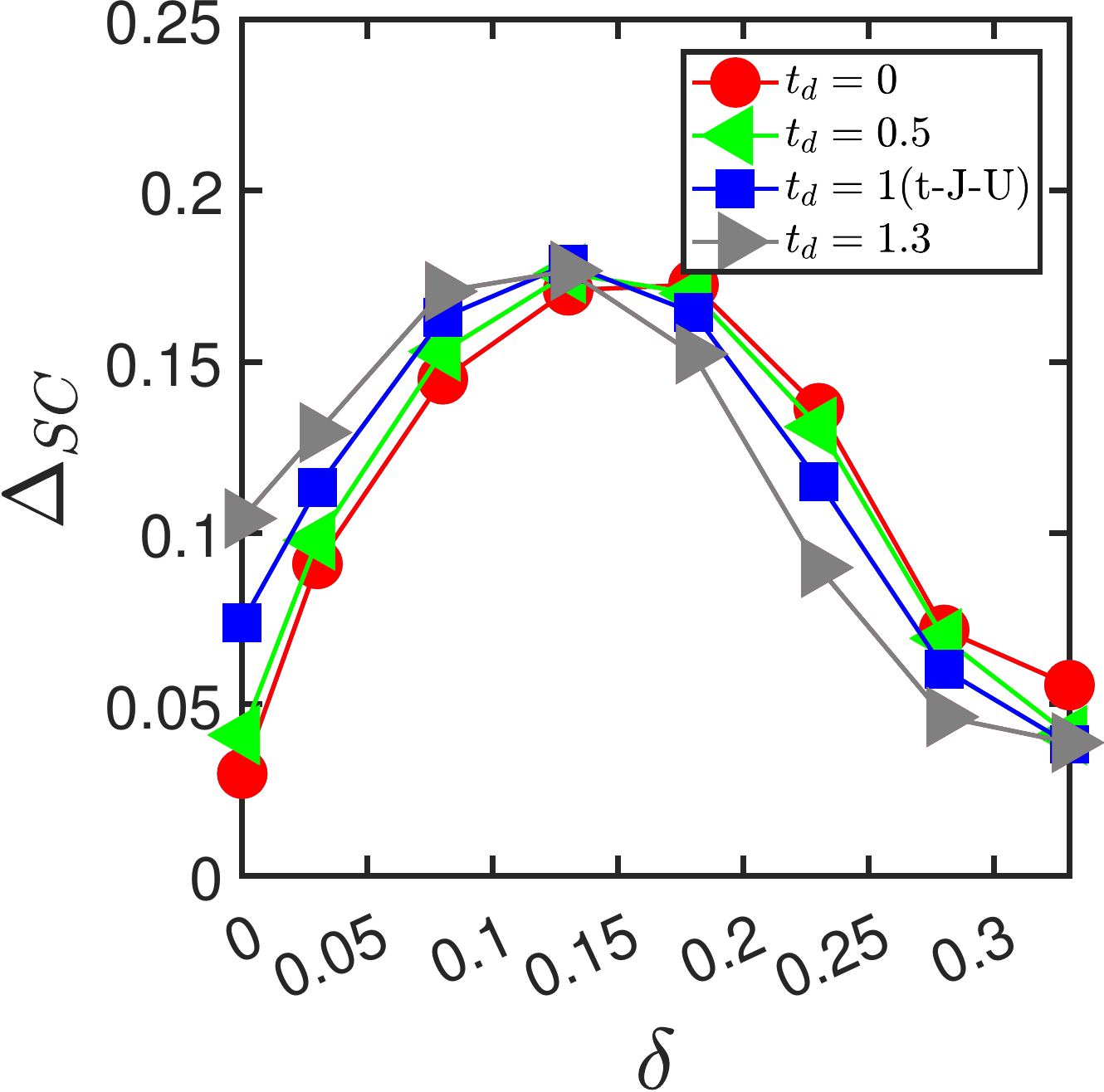}};
%		 	 {./Resulteffecttune/1Ub8tb1tp0OP}};
		\end{tikzpicture}
		\caption{}
		\label{sfig:efftd2}	
	\end{subfigure}
	\caption{Pairing order parameter correlation as a function of dopant. The CTG is set to (a)$\Delta_{CT} = 6$ and  (b)$\Delta_{CT} = 8$. The hopping parameter is $t = \bar t = 1 ~ J = 0.33$. }
	\label{fig:efftd}	%Second nearest neighbor $t' = t_d' = \bar t' = 0$.
\end{figure}

\begin{figure}[H]
	\centering
	\begin{subfigure}[b]{0.45\textwidth}
	\begin{tikzpicture}
	\node[inner sep=1] (image) {\includegraphics[width=\linewidth]{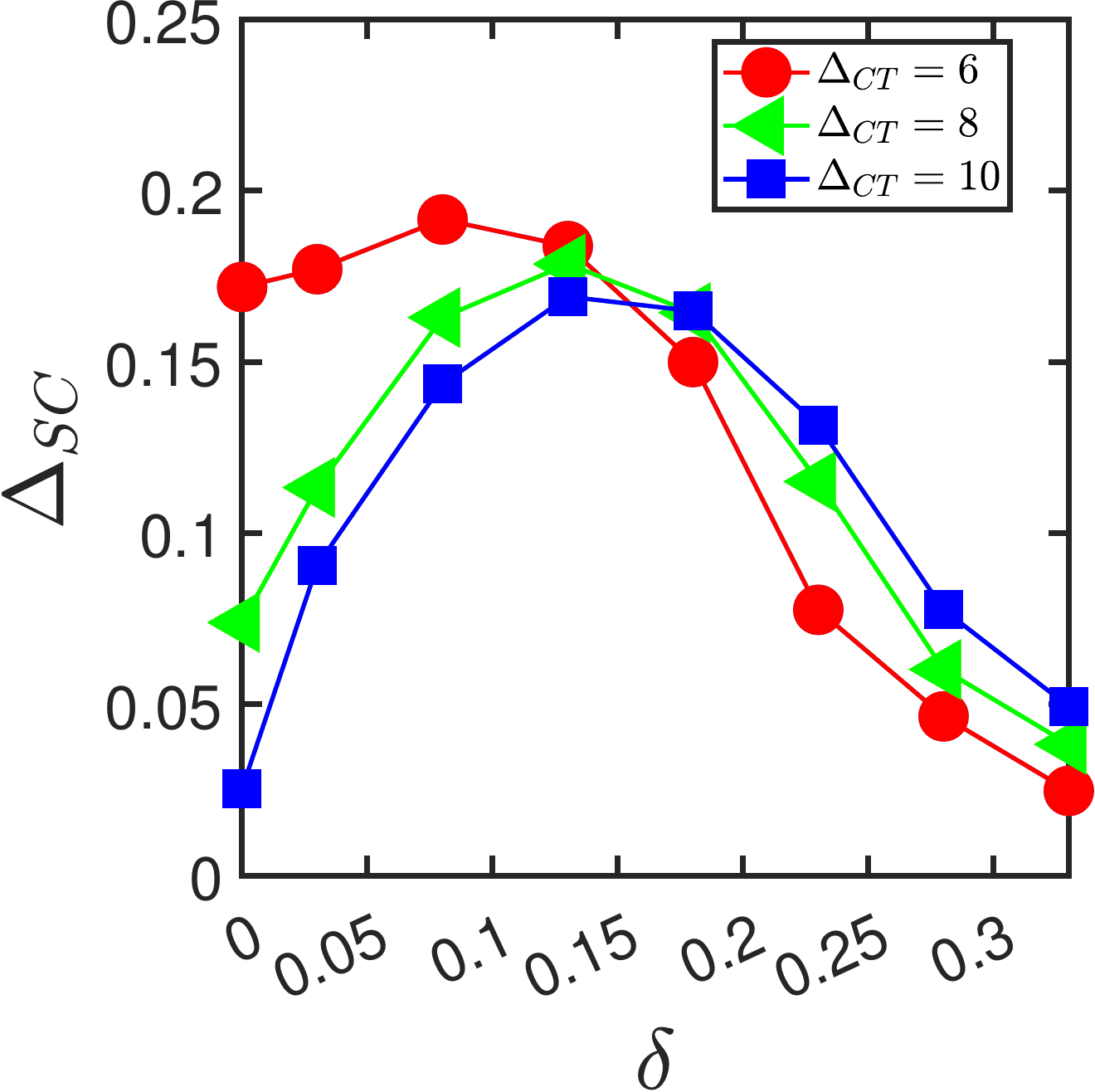}};
	%\begin{scope}[x={(image.south east)},y={(image.north west)}]
	%\draw[red,ultra thick,rounded corners] (0.65,0.10) rectangle (1,0.20);
	%\end{scope}
	\end{tikzpicture}
	\caption{}
	\label{sfig:effUbar1}	
\end{subfigure}
%\vspace{1em}
\begin{subfigure}[b]{0.45\textwidth}
	\begin{tikzpicture}
	%\node[anchor=south west,inner sep=0] (image) {\includegraphics[width=\linewidth] {./paper_D1/0RMFT_ppcd}};
	\node[anchor=south west,inner sep=1] (image) {\includegraphics[width=\linewidth] 		 	{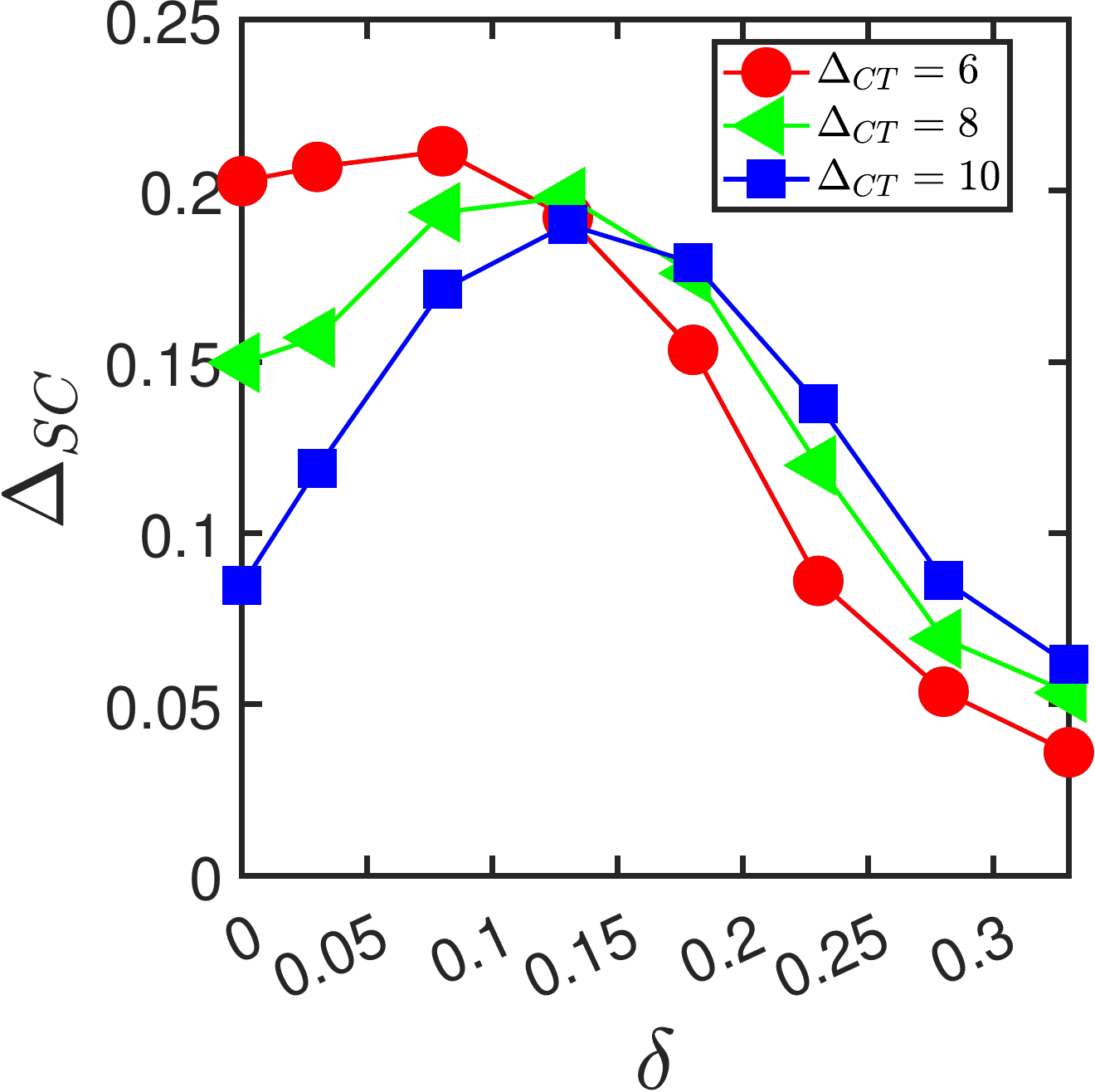}};
		%{./Resulteffecttune/0td1tb1.5tp0OP__}};
	%\begin{scope}[x={(image.south east)},y={(image.north west)}]
	%\draw[red,ultra thick,rounded corners] (0.65,0.08) rectangle (1,0.18);
	%\end{scope}
	\end{tikzpicture}
	\caption{}
	\label{sfig:effUbar2}	
\end{subfigure}
	
	\caption{Pairing order parameter as a function of dopant density with three different values of CTG. The hopping parameter is $t_d = 1$ and (a) is for $\bar t = 1$ and (b) is for $\bar t = 1.5$.
	}
	\label{fig:effUbar}	
\end{figure}

~~~~~~~In this section, we show
the $d$-wave pairing order parameter as a function of dopant concentration with varying $\bar t$, $t_d$ or $\Delta_{CT}$ in Figs. \ref{fig:efftbar}, \ref{fig:efftd} and \ref{fig:effUbar} respectively.

First, we consider two cases with $\Delta_{CT} = 6$ and $\Delta_{CT}=8$ in Figs \ref{sfig:efftbar1} and \ref{sfig:efftbar2}, respectively. By fixing $t_d = 1$, the CF process is only related to the formation of holon-doublon pair as shown in Fig. \ref{sfig:tbarhop1band}. 
Three hopping amplitudes $\bar t = 0.4$, $1$ and $1.5$ has been shown in the figure. Note that the label $\bar t = 0$ corresponds to  the $t-J$ model, while for $\bar t = 1$, it becomes the $t-J-U$ model.
Clearly, pairing is enhanced strongly in the underdoped(UD) regime as $\bar t$ increases. 
This enhancement of pairing can be understood by the increase in the effective superexchange interaction arises from the additional $\bar{t}$ process. When there is a larger $\Delta_{CT}$ and a smaller $\bar t$, the CF is less important and the result is closer to that of the $t-J$ model. Thus  CT is effectively reducing the Hubbard $U$. We also note that when $\bar t$ gets larger there is apparent finite superconductivity at very low doping, which is known as the Gossamer superconductivity\cite{Laughlin2002,PhysRevLett.90.207002,PhysRevB.71.014508}.

In Fig.~\ref{fig:efftd}, we study the effect of $t_d$. The $d$-wave pairing order parameter as a function of dopant concentration for  $t_d = 0$, $0.5$, $1$ and $1.3$ is shown in Fig.~\ref{sfig:efftd1} and ~\ref{sfig:efftd2}
for $\Delta_{CT} = 6$ and $\Delta_{CT}=8$, respectively. Notice here we set that $\bar t = 1$. 
By increasing $t_d$, we found  pairing at UD regime is slightly enhanced but suppressed at the OD. Similar to the effect of  $\bar t$, increasing $t_d$ enhances CF and suppresses strong correlation. The suppression of pairing in the OD regime is due to the lack of 
singly occupied spin-$1/2$ particles or spinons,  which effectively reduces the number of electrons available for pairing. The controlling parameter of spinon density and how it affects pairing will be elaborated further in the next section.

In Fig~\ref{fig:effUbar}, we show the $d$-wave pairing order parameter as a function of dopant concentration for $t_d = 1$ and three different CTG, $\Delta_{CT}=6$, $8$ and $10$. Fig. \ref{sfig:effUbar1} is for $\bar t = 1$ and \ref{sfig:effUbar2} for 1.5.
Similar to the result of large $t_d$ or $\bar t$, smaller $\Delta_{CT}$ enhances pairing in the UD regime but suppress it in the OD side.

All the results shown in Figs. 2-4 reveal a consistent physical picture. When $\Delta_{CT}$ 
becomes smaller or $\bar t$ or $t_d$ becomes larger, maximum SC order parameter increases and the pairing dome moves toward 
the smaller dopant concentration. In addition we found qualitatively similar result with the $t-J-U$ model for various 
values of $\bar t$ and $t_d$. On the other hand, the \enquote{shift} of pairing dome in Fig.~\ref{fig:effUbar} is very similar to the high pressure experiment result on Hg1212 \cite{Yamamoto2015}. This will be discussed further in Sec.\ref{sec:xxx}
%Hence from now on, we will often use $t-J-U$ model to discuss the physics revealed by our results except when we discuss the pressure effect.
%The results of including second neighboring hopping $t'/t =t_d^{\prime}/t_d=\bar t'/\bar t = -0.3$ are shown in supplementary material which have very similar with the result of $t'=0$.

%The numerical result indicates that moving toward strong CF or effective weak Hubbard U(which is $\Delta_{CT}$ in our model) enhances maximum pairing amplitude, which agrees with experiment result\cite{Ruan2016}. 

\subsection{Relation of CTG and pairing amplitude } \label{sec:ctgtc}
%\maketitle{How charge transfer gap influence $T_c$}

\begin{figure}[H]
	\centering
	\begin{tikzpicture}
	\node[inner sep=1] (image) {\includegraphics[width=0.7\linewidth] {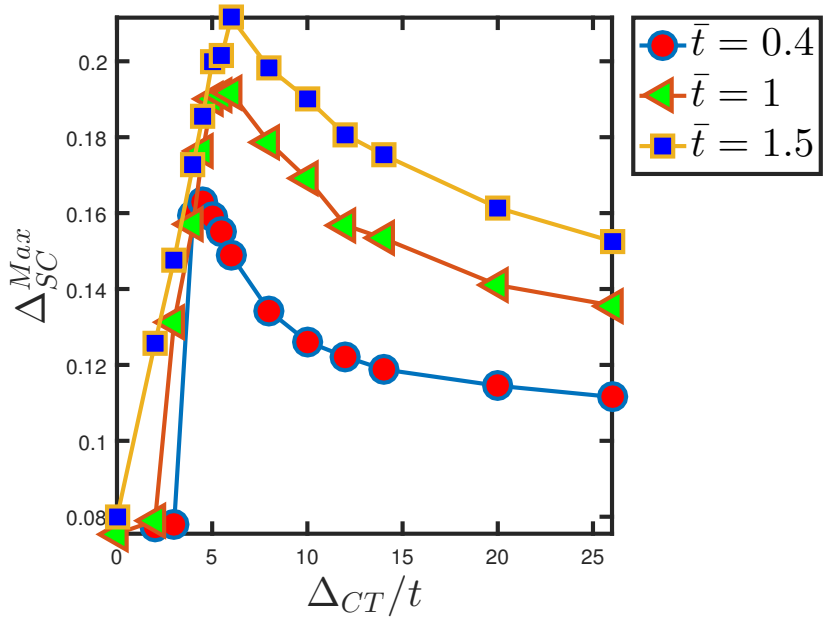}};;
	%\begin{scope}[x={(image.south east)},y={(image.north west)}]
	%\draw[red,ultra thick,rounded corners] (0.65,0.10) rectangle (1,0.20);
	%\end{scope}
	\end{tikzpicture}
	\caption{Maximum $d$-wave pairing order parameter vs. CTG with $t_d = 1$ and $t' = 0$. 
		%The $\Delta_{SC}^{Max}$ is fitted by the empirical formula $(1 - \frac{T_c}{T_c^{max}}) = 82.6(\delta - \delta_{opt})^2$\cite{PRESLAND199195,J.L.TallonC.BernhardH.ShakedR.L.Hitterman1995}. 
	}	
	\label{fig:pairMax}
\end{figure}

\begin{figure}[H]
 	\centering
 	\begin{subfigure}{0.7\linewidth}\label{sfig:ns}
		%\begin{tikzpicture}	
		%\node[anchor=south west,inner sep=1] (image)
		\centering
		{\includegraphics[width=\textwidth] {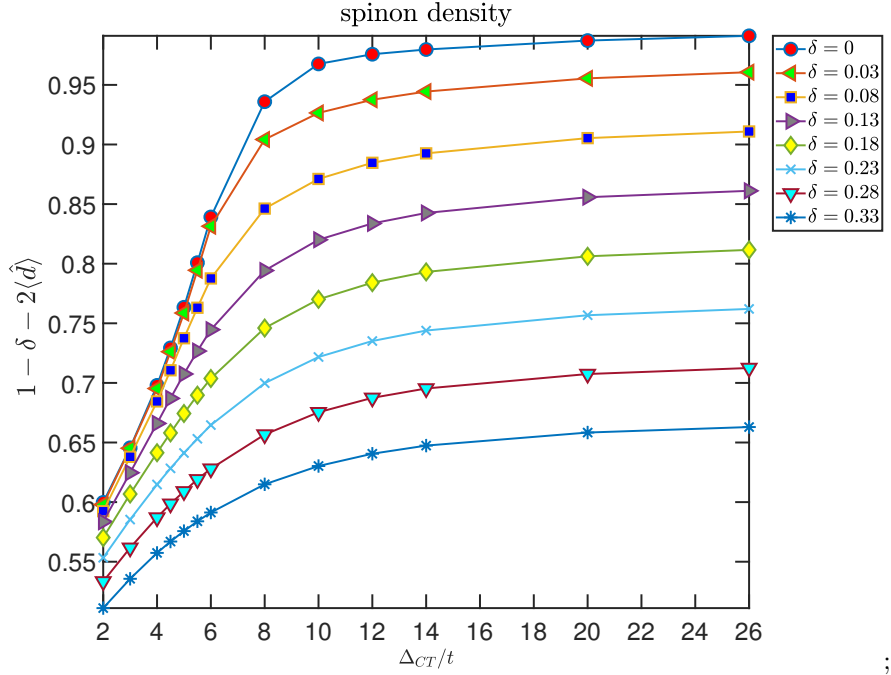}};
		%\end{tikzpicture}		
 	\end{subfigure}
 	\hfill
 	\begin{subfigure}{0.7\linewidth}		\label{sfig:ppcdu}		
		%\begin{tikzpicture}
		%\node[inner sep=1] (image)
		\centering
		 {\includegraphics[width=\textwidth] 	{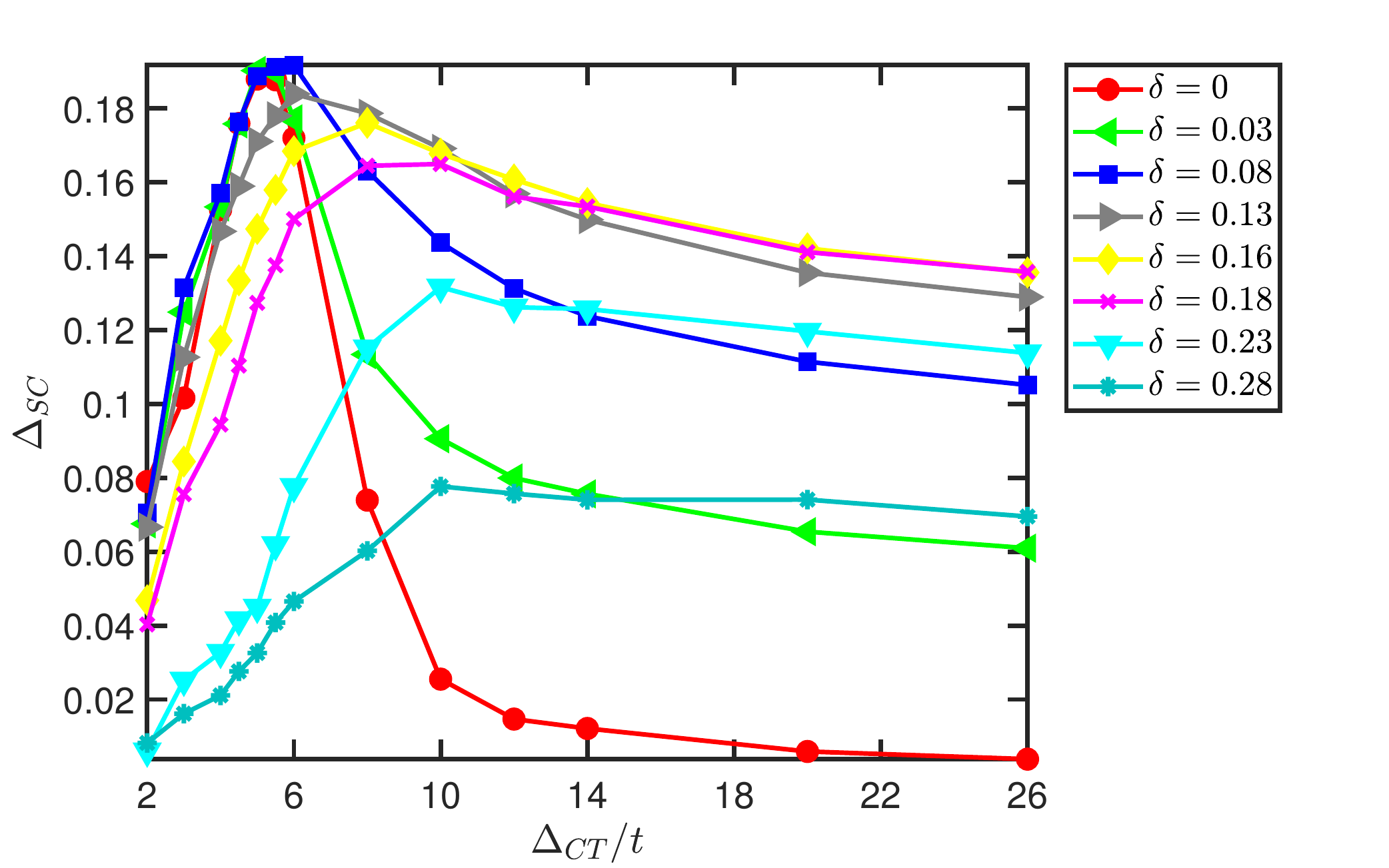}};;
		%\end{tikzpicture} 		
 	\end{subfigure} 	
	\caption{ (a)spinon density as a function of $\Delta_{CT}$.
	(b)Pairing order parameter as a function of $\Delta_{CT}$.
	The parameters are $\bar t = t_d  = 1$ and $t'=0$.}
\label{fig:ppcdxx}
 \end{figure}

~~~~~~Recently, Ruan $\it et al.$ \cite{Ruan2016} found anti-correlation between $T_c^{Max}$ and CTG in both single- and double- layered cuprate via scanning tunneling
 spectroscopy(STS).  To investigate the anti-correlation, we 
 determine the maximum pairing amplitude, $\Delta_{SC}^{Max}$, as a function of the CTG by calculating the paring amplitude as a function of hole density. In Fig. \ref{fig:pairMax}, for small $\Delta_{CT}$ or effective U, where Gossamer  superconductivity\cite{Laughlin2002,PhysRevLett.90.207002,PhysRevB.71.014508} dominates, $\Delta_{SC}^{Max}$ is approximately proportional to $\Delta_{CT}$ with almost the same slope. This can be explained simply.  As $\Delta_{CT}$ increases from 0, doublon number begins to decrease significantly and consequently the number of singly occupied sites or spinon increases (see Fig. \ref{fig:ppcdxx}a).
  Once the CTG values reach a critical magnitude, the strong correlation Mott physics begins to take hold, then the increase of $\Delta_{CT}$ will result in a decrease of pairing order which should be inversely proportional to the CTG in the form of $t^2/\Delta_{CT}$.
  This is consistent with the result in Ref. \cite{Ruan2016} that  $T_c^{max}$ is roughly proportional to $1/\Delta_{CT}$ for one- and two-layer cuprates.  Their data also shows when the CTG is reduced by 25$\%$  from one layer CCOC to Bi2201, the $T_c^{max}$ increases 45$\%$. By choosing $t=0.3eV$ and consider $\bar t=0.4t$, we get the enhancement of maximum pairing amplitude about $23\%$ when $\Delta_{CT}/t$ is reduced from $6.5$ to $5$. This result is quite satisfactory considering that  the parameters of these two  different cuprates are roughly estimated. 
\begin{comment}
\begin{figure}
	\centering
%	\begin{subfigure}[b]{0.5\textwidth}
		\begin{tikzpicture}
		\node[inner sep=1] (image) {\includegraphics[width=0.7\linewidth] {./paper_D3.5/L20J0.33tbar1t_d1dopant0.13L20J0.33tbar1t_d1dopant0.13}};;
		%\begin{scope}[x={(image.south east)},y={(image.north west)}]
		%\draw[red,ultra thick,rounded corners] (0.65,0.10) rectangle (1,0.20);
		%\end{scope}
		\end{tikzpicture}
%		\label{sfig:reproDMRG1}	
%	\end{subfigure}
	~
	%	\begin{subfigure}[b]{0.7\textwidth}
	%		\begin{tikzpicture}
	%\node[anchor=south west,inner sep=0] (image) {\includegraphics[width=\linewidth] {./paper_D1/0RMFT_ppcd}};
	%	\node[inner sep=1] (image) {\includegraphics[width=\linewidth] {./paper_D3.5/dos_0.13}};;
	%\begin{scope}[x={(image.south east)},y={(image.north west)}]
	%\draw[red,ultra thick,rounded corners] (0.65,0.08) rectangle (1,0.18);
	%\end{scope}
	%		\end{tikzpicture}
	%		\caption{}
	%		\label{sfig:reproDMRG2}	
	%	\end{subfigure}
	\caption{
		(a)Pairing order parameter as a function of CTG $\Delta_{CT}$ with hole density $ ~\delta = 0.13$. The hopping parameter is set as $ \bar t  = t_d=1$ and $t' = 0$. 
This is same as the familiar $t-J-U$ model. 
		%(b) The density of state near the Fermi level vs. CTG. The density of state is calculated by non-interacting energy dispersion $\xi_k$ with the optimized variational parameter $t^{\prime}_v$ and $\mu_v$ as mentioned in section.\ref{sec:Formalism}. 
	}	
	\label{fig:reproDMRG}
\end{figure}
\end{comment}

Since the introduction of the CTG provides an additional superexchange interaction for pairing, this sure will enhance $\Delta_{SC}^{Max}$ and  $T_c^{Max}$.
However, we were surprised to find  that pairing order parameter is not uniformly enhanced for all the doped hole density when CTG is reduced as shown in Fig. \ref{fig:effUbar}. The enhancement is larger in the UD regime and with no enhancement or pairing is even suppressed in the OD phase. As noted by Anderson\cite{ANDERSON1196} a long time ago, the $d$-wave pairing caused by the superexchange interaction favors the singly occupied sites or spinons. In Fig~\ref{fig:ppcdxx}a, we examine the effect of spinon density on pairing. The  parameters used are $\bar t = t_d  = t$, $J = 0.33$ and $t' = t_d' = \bar t' = 0$  so this is same as $t-J-U$ model.  For all the concentrations, the spinon density first increases quickly with CTG  and then approaches a plateau with a very small increase  around $\Delta_{CT}=8-10$ which enters the strong correlation regime.
The increase of spinon density is much rapid and larger in the UD regime than in OD regime. Thus we can expect pairing will be quickly enhanced in the UD regime but much milder in the OD regime. Accordingly, one would also expect $\Delta_{SC}^{Max}$ to be larger in the UD regime. This is exactly what we found in Fig.~\ref{fig:ppcdxx}b, where the pairing order parameter is plotted as a function of CTG for various hole densities. For hole density less than $~0.18$, the pairing order rapidly increases with CTG. Once the maximum value is reached at some critical CTG, the system enters the strongly correlated regime and  
the increase of $\Delta_{CT}$ only reduces the additional superexchange interaction thus pairing order decreases as $1/\Delta_{CT}$. At very large values of $\Delta_{CT}$, the system is essentially the same as the $t-J$ model and the pairing order is quite small as it is proportional to the hole density in the UD regime.  In the OD regime, although the spinon density still gradually increases as CTG is increased, it is weaker than the effect of decreases of additional superexchange interaction as $1/ \Delta_{CT}$.
% In the OD regime, the pairing also increases with increase of CTG but much more gradually due to the small increase of spinon density shown in Fig.~\ref{fig:ns}. 
% Interestingly, we notice that $\Delta_{SC}^{Max}$ is indeed larger with a value about 0.19 in the UD regime than OD. 

%  On the other hand, if we set $J=0$, the model will become Hubbard model. Similar result for Hubbard model can be found in Ref.\cite{Yokoyama2013a}.
  
\subsection{Examine the effect of pressure} \label{sec:xxx}

\begin{figure}[H]
	\centering
	\begin{subfigure}[b]{0.45\textwidth}
		\begin{tikzpicture}
		\node[inner sep=1] (image) {\includegraphics[width=\linewidth]{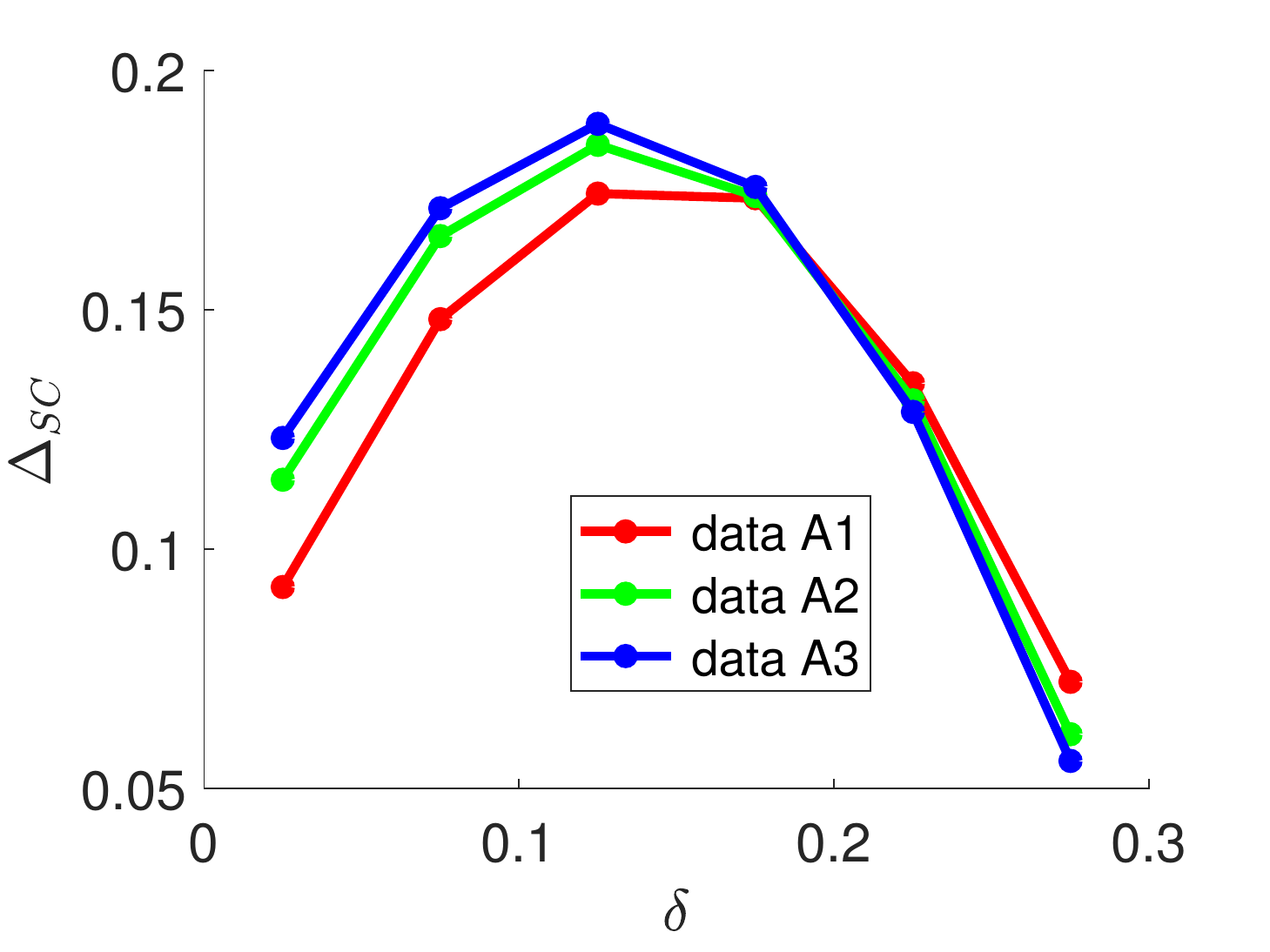}};
		\end{tikzpicture}
		\caption{}
		\label{sfig:p0}	
	\end{subfigure}
%\hspace{1em}
	\begin{subfigure}[b]{0.45\textwidth}
		\begin{tikzpicture}
		\node[anchor=south west,inner sep=1] (image) {\includegraphics[width=\linewidth] 		 	{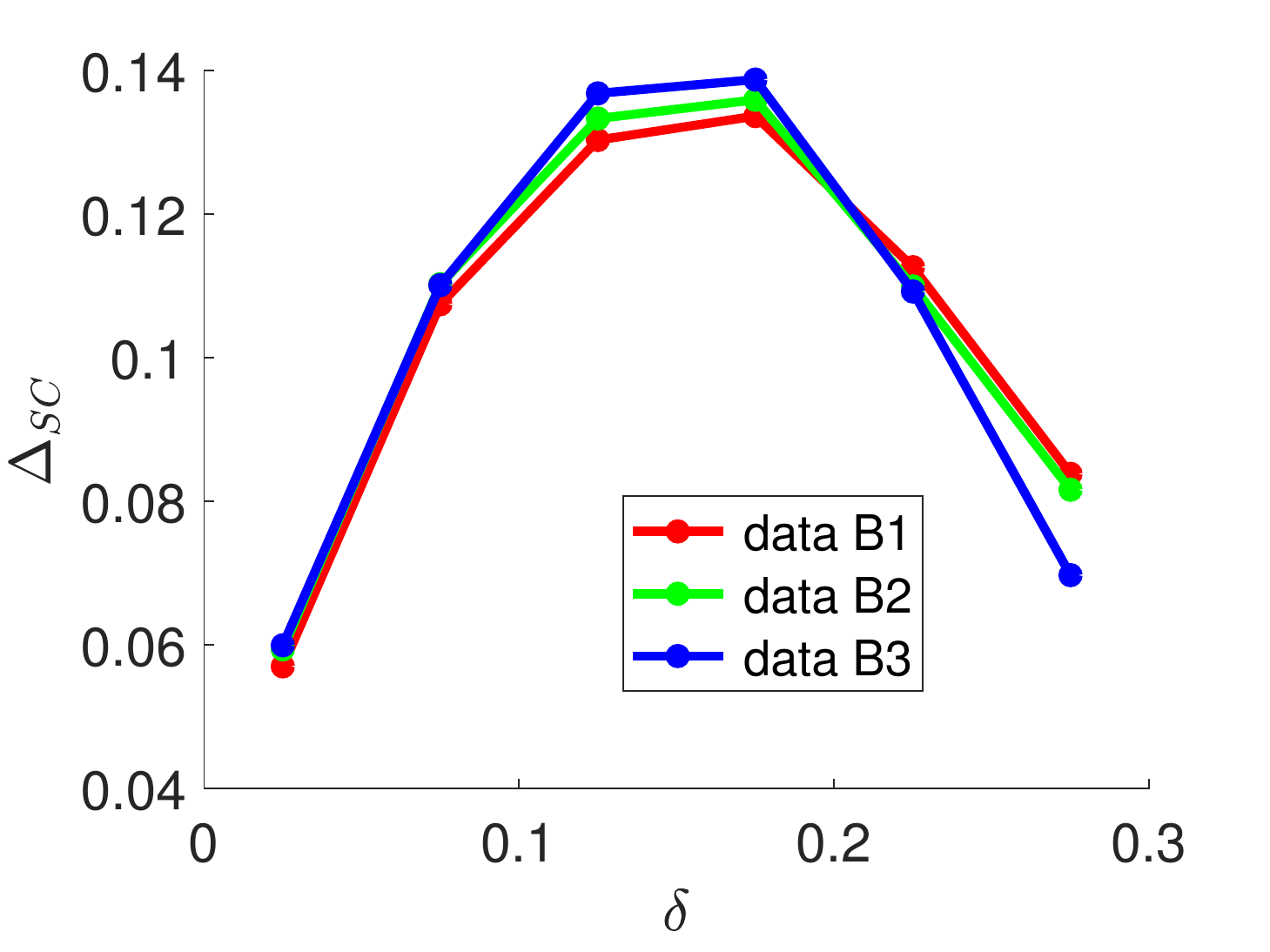}};
		\end{tikzpicture}
		\caption{}
		\label{sfig:p1}	
	\end{subfigure}	
 	\caption{Simulation of pressure effect, parameter at ambient pressure($J/t$, $\Delta_{CT}/t$, $t_d/t$, $\bar t_/t$) are seleted to be (0.33, 8, 1, 1) for (a) and
	(0.33, 8, 0.5, 0.4) for (b). 		
	The detail parameters are listed in Table.\ref{tab:p0}}	
	\label{fig:p}	
\end{figure}

When pressure is applied, the in-plane distance between Cu and Oxygen will decrease, hence $t_{pd}$ will increase.
As shown from previous sections, not only $\Delta_{CT}$ but also $t_d$ and $\bar t$ can affect the superconducting dome shape. According to standard perturbation derivation: $t \sim t_{pd}^2$, $\bar t \sim t_{pd}$, $t_d \sim t_{pd}^2$, $J \sim t_{pd}^4$. Hence as $t_{pd}$ increases, $t_d/t$  hardly changes, $\bar t/t$ is inversely proportional to $t_{pd}$, thus it will decreases,  $\Delta_{CT}/t$ will decreases as $1/{t_{pd}^2}$. and $J/t$ is proportional to $t_{pd}^2$. So assuming that we set the parameter at ambient pressure($J/t$, $\Delta_{CT}/t$, $t_d/t$, $\bar t_/t$) to be (0.33, 8, 1, 1).
If we increase $t_{pd}$ by a ratio $\alpha$ then the parameters will turn into
($0.33\alpha^2, 8/\alpha^2, 1, 1/\alpha$).
As pressure is increased, $\alpha$ increases from 1 to 1.04 and 1.08 as shown in Table \ref{tab:p0} with all other parameters.
 In Fig.\ref{fig:p}, the two sets of parameters show the similar result that as pressure or $t_{pd}$  increases, pairing order parameter increases in UD regime  but decreases in OD.
 This is very consistent with  high pressure experiment\cite{Yamamoto2015}.

\begin{table}[H]
	\centering
	\caption{}
 	\label{tab:p0}
	\begin{tabular}{|l|l|l|l|l|l|l|l|l|l|}
		\hline
		data label       & A1    &A2     & A3  & B1    &B2     & B3  \\ \hline
		$\alpha$ 		 & 1     & 1.04 & 1.08 & 1     & 1.04 & 1.08  \\ \hline
		$J/t$            & 0.33 & 0.357 & 0.385& 0.33 & 0.357 & 0.385 \\ \hline
		$\Delta_{CT}/t$  & 8    & 7.4   & 6.86 & 8    & 7.4   & 6.86\\ \hline
		$t_d/t$          & 1  & 1   & 1 & 0.5  & 0.5   & 0.5\\ \hline
		$\bar t /t$      & 1  & 0.961 & 0.926& 0.4  & 0.384 & 0.37 \\ \hline
	\end{tabular}
\end{table}

\begin{figure}[H]
	\centering
	\begin{subfigure}[b]{0.45\textwidth}
		\begin{tikzpicture}
		\node[inner sep=1] (image) {\includegraphics[width=\linewidth]{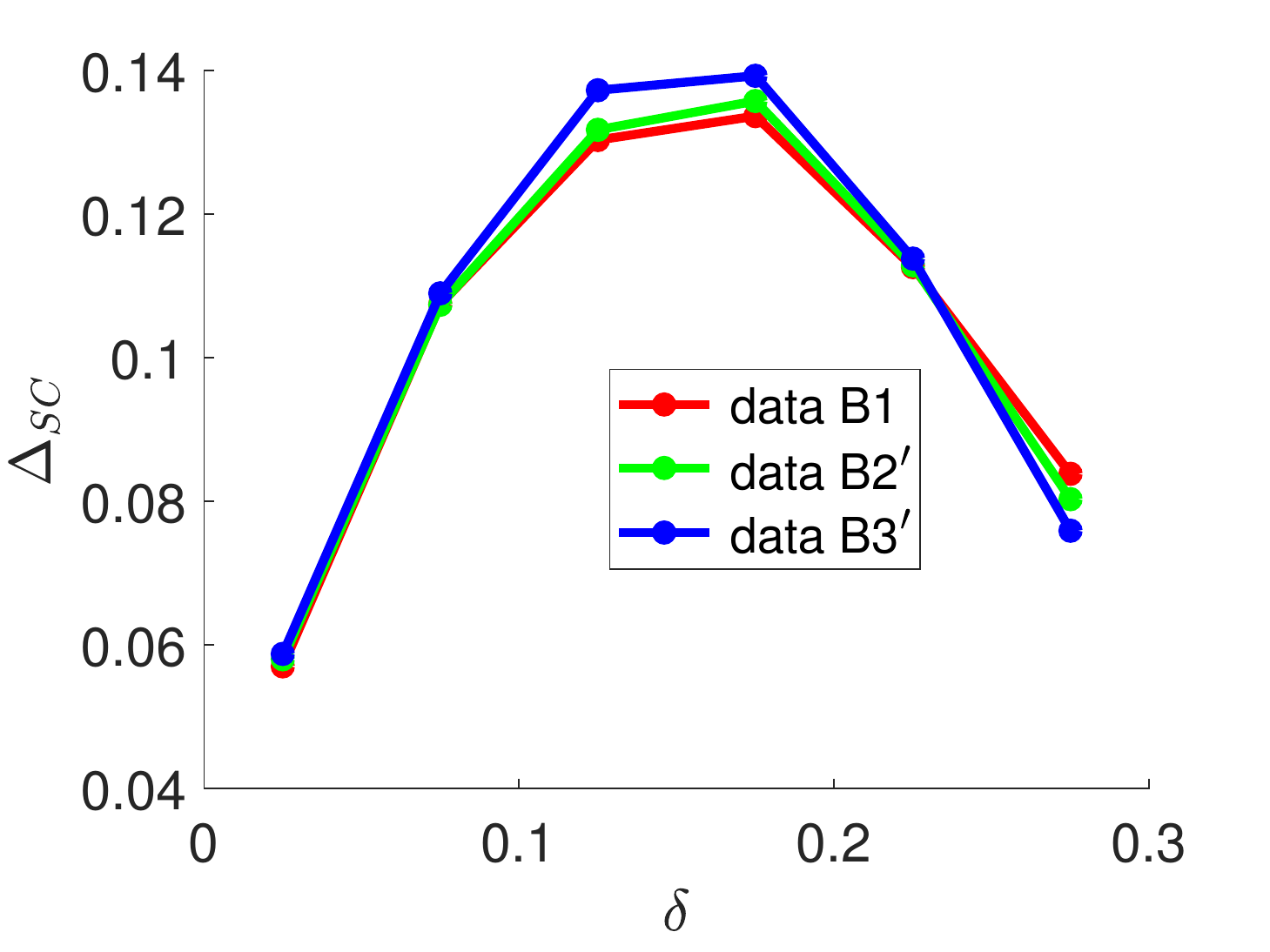}};
		\end{tikzpicture}
		\caption{}
		\label{sfig:U025}	
	\end{subfigure}
	%\hspace{1em}
	\begin{subfigure}[b]{0.45\textwidth}
		\begin{tikzpicture}
		\node[anchor=south west,inner sep=1] (image) {\includegraphics[width=\linewidth] 		 	{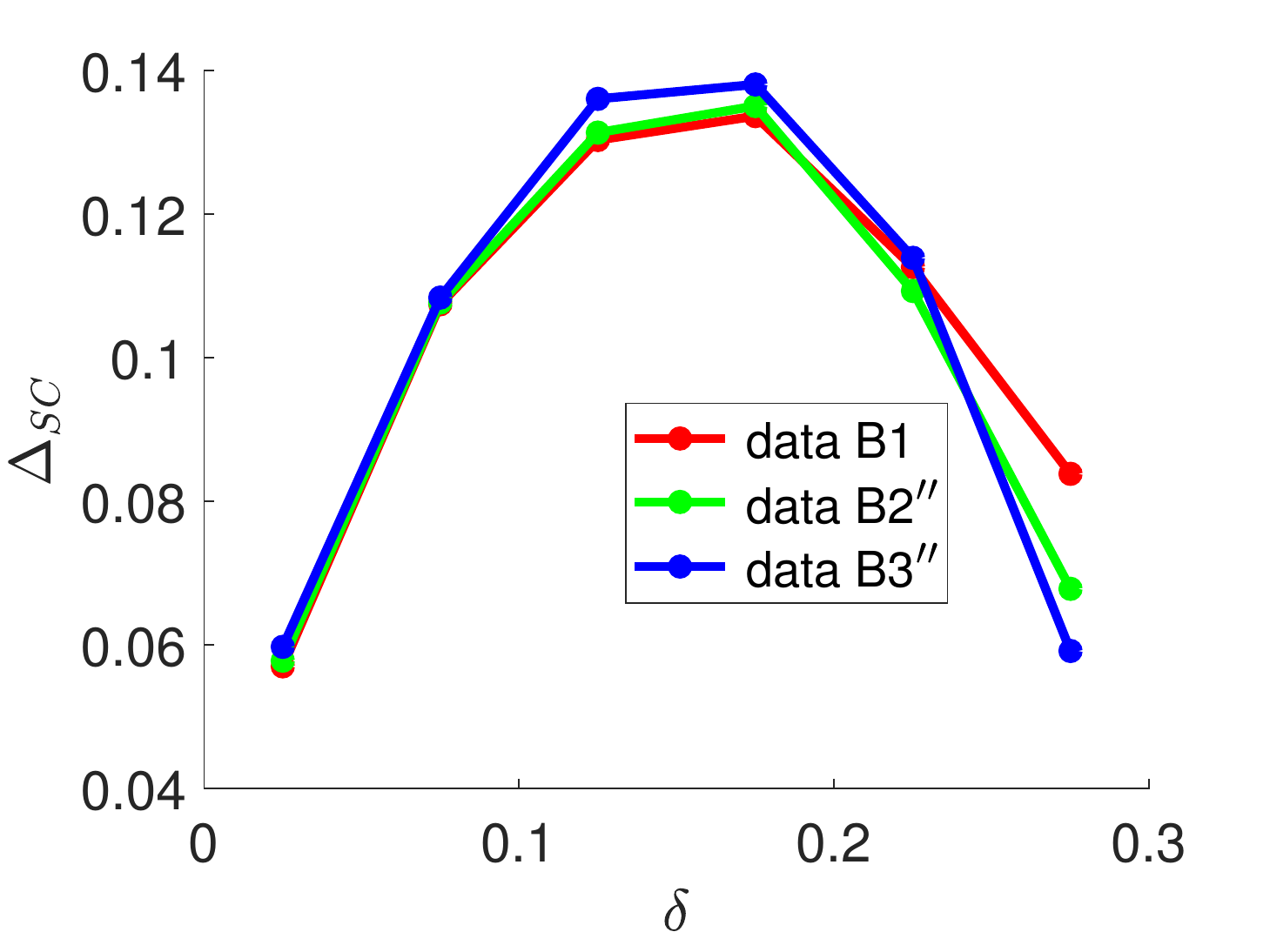}};
		\end{tikzpicture}
		\caption{}
		\label{sfig:U05}	
	\end{subfigure}	
	\caption{Simulation of pressure effect that include the apical oxygen effect, parameter at ambient pressure($J/t$, $\Delta_{CT}/t$, $t_d/t$, $\bar t_/t$) are seleted to be (0.33, 8, 0.5, 0.4). As pressure is applied, the parameters becomes (0.33$\alpha$, 8$\beta$/$\alpha^2$, 0.5, 0.4/$\alpha$)The detail parameter for the data is listed in Table.\ref{tab:U}}	
	\label{fig:U}	
\end{figure}

On the other hand, the hydrostatic pressure will also decrease the apical oxygen distance. When the apical Oxygen is brought closer to the Cu atom, it will increase the repulsion to electrons on the copper, hence the increase of charge transfer gap\cite{Weber2010a,Weber2010}.  
distance.  To simulate this effect, we increase $\Delta_{CT}$  a little bit, say $2.5\%$ or $5\%$.
Assuming at ambient pressure the parameter ($J/t$, $\Delta_{CT}/t$, $t_d/t$, $\bar t_/t$) is (0.33, 8, 0.5, 0.4) as data of B1 in Table \ref{tab:p0}, With increasing pressure, the $t_{pd}$ becomes $\alpha t_{pd}$ and 
$\Delta_{CT}$ becomes $\beta \Delta_{CT}$. Hence, as pressure is applied, the parameters becomes (0.33$\alpha^2$, 8$\beta$/$\alpha^2$, 0.5, 0.4/$\alpha$).
The detailed parameters are listed in Table.\ref{tab:U}. Fig. \ref{fig:U} shows similar result as Fig. \ref{fig:p} although we have included the effect that CTG is increased by reducing the apical oxygen distance from Cu. The UD side has its pairing enhanced while the OD side is reduced under pressure as the pairing dome is shifted or tilted toward the UD side.

\begin{table}[H]
	\centering
	\caption{}
	\label{tab:U}
	\begin{tabular}{|l|l|l|l|l|l|l|l|l|l|}
		\hline
		data label       & B1    & $B2^{\prime}$     & $B3^{\prime}$  & $B2^{\prime\prime}$    & $B3^{\prime\prime}$ \\ \hline
		$\alpha$ 		 & 1     & 1.04 & 1.08  & 1.04 & 1.08  \\ \hline
		$\beta$ 		 & 1     & 1.025 & 1.025 & 1.05     & 1.05 \\ \hline
		$J/t$            & 0.33 & 0.357 & 0.385& 0.357 & 0.385 \\ \hline
		$\Delta_{CT}/t$  & 8    & 7.58   & 7.03 & 7.77    & 7.2  \\ \hline
		$t_d/t$          & 0.5  & 0.5   & 0.5 & 0.5  & 0.5  \\ \hline
		$\bar t /t$      & 0.4  & 0.384 & 0.37& 0.384  & 0.37 \\ \hline
	\end{tabular}
\end{table}

	\section{Conclusion}\label{conclusion}	

~~~~~~~In summary, motivated by the recent experiments \cite{Cai2016}, we have studied the CF effect of the Copper-Oxygen three-band model with an effective one-band model similar to the t-J-U model. But now U or $\Delta_{CT}$ is the charge transfer gap. The formation of this doublon and its hopping as well as the size of the charge transfer gap all played an important role in the CF effect.  The inclusion of these effects provides a simple understanding of the enhancement of maximum pairing when the $\Delta_{CT}$ is reduced as observed by the experiment \cite{Ruan2016}. Additionally, we also found that there is a minimum $\Delta_{CT}$ for enhancement.  On the other hand, this enhancement only occurs in the UD regime but not in OD. This dichotomy between UD and OD is further examined  by considering the effect of applying pressure. Our model provides a very simple explanation for the observations in several pressure experiments that $T_c$ is always enhanced in the UD regime but reduced in the OD.
%as the $\Delta_{CT}$ is reduced is consistent with the applying pressure to cuprates. $T_c$ is enhanced in the UD or optimal doping but reduced in the OD regime.  

%Also, we proved that tuning the $t_{pd}$ can cause the "crossing" of superconducting dome which can be used to simulate the effect of applying hydrostatic pressure.
%To have a more qualitative comparison with experiments, we 
% acknowledgement.tex                            19 February 2016
%
% Acknowledgement paragraph with agency abbreviations of Oct. 15, 2014 for non-APS journals

\section* {\textbf{Acknowledgement}}
YHL and TKL were supported by Taiwan Ministry of Science and Technology Grant 108-2112-M-110-015. 
The calculation was supported by Academia Sinica Grid-computing Center(ASGC) and National Center for High Performance Computing in Taiwan.   % input acknowledgement
%\section*{References}

%\bibliographystyle{unsrt}
%\addcontentsline{toc}{chapter}{\bibname}
\bibliography{./main}

\begin{thebibliography}{10}
\expandafter\ifx\csname url\endcsname\relax
  \def\url#1{\texttt{#1}}\fi
\expandafter\ifx\csname urlprefix\endcsname\relax\def\urlprefix{URL }\fi
\expandafter\ifx\csname href\endcsname\relax
  \def\href#1#2{#2} \def\path#1{#1}\fi

\bibitem{Bednorz1986}
J.~G. Bednorz, K.~A. M{\"u}ller,
  \href{http://dx.doi.org/10.1007/BF01303701}{{Possible highTc
  superconductivity in the $Ba−La−Cu−O$ system}}, Zeitschrift f{\"u}r
  Physik B Condensed Matter 64~(2) (1986) 189.
\newblock \href {https://doi.org/10.1007/BF01303701}
  {\path{doi:10.1007/BF01303701}}.
\newline\urlprefix\url{http://dx.doi.org/10.1007/BF01303701}

\bibitem{PhysRevLett.58.908}
M.~K. Wu, J.~R. Ashburn, C.~J. Torng, P.~H. Hor, R.~L. Meng, L.~Gao, Z.~J.
  Huang, Y.~Q. Wang, C.~W. Chu,
  \href{https://link.aps.org/doi/10.1103/PhysRevLett.58.908}{{Superconductivity
  at 93 K in a new mixed-phase Y-Ba-Cu-O compound system at ambient pressure}},
  Phys. Rev. Lett. 58 (1987) 908.
\newblock \href {https://doi.org/10.1103/PhysRevLett.58.908}
  {\path{doi:10.1103/PhysRevLett.58.908}}.
\newline\urlprefix\url{https://link.aps.org/doi/10.1103/PhysRevLett.58.908}

\bibitem{RevModPhys.78.17}
P.~A. Lee, N.~Nagaosa, X.-G. Wen,
  \href{https://link.aps.org/doi/10.1103/RevModPhys.78.17}{Doping a mott
  insulator: Physics of high-temperature superconductivity}, Rev. Mod. Phys. 78
  (2006) 17--85.
\newblock \href {https://doi.org/10.1103/RevModPhys.78.17}
  {\path{doi:10.1103/RevModPhys.78.17}}.
\newline\urlprefix\url{https://link.aps.org/doi/10.1103/RevModPhys.78.17}

\bibitem{doi:10.1080/00018730701627707}
B.~Edegger, V.~N. Muthukumar, C.~Gros,
  \href{https://doi.org/10.1080/00018730701627707}{{Gutzwiller–RVB theory of
  high-temperature superconductivity: Results from renormalized mean-field
  theory and variational Monte Carlo calculations}}, Advances in Physics 56~(6)
  (2007) 927--1033.
\newblock \href {https://doi.org/10.1080/00018730701627707}
  {\path{doi:10.1080/00018730701627707}}.
\newline\urlprefix\url{https://doi.org/10.1080/00018730701627707}

\bibitem{Ogata2008}
M.~Ogata, H.~Fukuyama,
  \href{http://stacks.iop.org/0034-4885/71/i=3/a=036501{\%}5Cnpapers3://publication/doi/10.1088/0034-4885/71/3/036501}{{The
  t -- J model for the oxide high- T c superconductors}}, Reports on Progress
  in Physics 71~(3) (2008) 36501.
\newblock \href {http://arxiv.org/abs/arXiv:1008.0522v1}
  {\path{arXiv:arXiv:1008.0522v1}}, \href
  {https://doi.org/10.1088/0034-4885/71/3/036501}
  {\path{doi:10.1088/0034-4885/71/3/036501}}.
\newline\urlprefix\url{http://stacks.iop.org/0034-4885/71/i=3/a=036501{\%}5Cnpapers3://publication/doi/10.1088/0034-4885/71/3/036501}

\bibitem{ANDERSON1196}
P.~W. ANDERSON, \href{http://science.sciencemag.org/content/235/4793/1196}{{The
  Resonating Valence Bond State in $La_2CuO_4$ and Superconductivity}}, Science
  235~(4793) (1987) 1196.
\newblock \href
  {http://arxiv.org/abs/http://science.sciencemag.org/content/235/4793/1196.full.pdf}
  {\path{arXiv:http://science.sciencemag.org/content/235/4793/1196.full.pdf}},
  \href {https://doi.org/10.1126/science.235.4793.1196}
  {\path{doi:10.1126/science.235.4793.1196}}.
\newline\urlprefix\url{http://science.sciencemag.org/content/235/4793/1196}

\bibitem{PhysRevLett.58.2794}
V.~J. Emery, \href{http://link.aps.org/doi/10.1103/PhysRevLett.58.2794}{{Theory
  of high-${\mathrm{T}}_{\mathrm{c}}$ superconductivity in oxides}}, Phys. Rev.
  Lett. 58 (1987) 2794.
\newblock \href {https://doi.org/10.1103/PhysRevLett.58.2794}
  {\path{doi:10.1103/PhysRevLett.58.2794}}.
\newline\urlprefix\url{http://link.aps.org/doi/10.1103/PhysRevLett.58.2794}

\bibitem{PhysRevB.37.3759}
F.~C. Zhang, T.~M. Rice,
  \href{http://link.aps.org/doi/10.1103/PhysRevB.37.3759}{{Effective
  Hamiltonian for the superconducting Cu oxides}}, Phys. Rev. B 37 (1988) 3759.
\newblock \href {https://doi.org/10.1103/PhysRevB.37.3759}
  {\path{doi:10.1103/PhysRevB.37.3759}}.
\newline\urlprefix\url{http://link.aps.org/doi/10.1103/PhysRevB.37.3759}

\bibitem{PhysRevB.41.7243}
F.~C. Zhang, T.~M. Rice,
  \href{http://link.aps.org/doi/10.1103/PhysRevB.41.7243}{Validity of the
  \textit{t} - \textit{J} model}, Phys. Rev. B 41 (1990) 7243.
\newblock \href {https://doi.org/10.1103/PhysRevB.41.7243}
  {\path{doi:10.1103/PhysRevB.41.7243}}.
\newline\urlprefix\url{http://link.aps.org/doi/10.1103/PhysRevB.41.7243}

\bibitem{PhysRevB.88.134525}
Y.-J. Chen, M.~G. Jiang, C.~W. Luo, J.-Y. Lin, K.~H. Wu, J.~M. Lee, J.~M. Chen,
  Y.~K. Kuo, J.~Y. Juang, C.-Y. Mou,
  \href{https://link.aps.org/doi/10.1103/PhysRevB.88.134525}{{Doping evolution
  of Zhang-Rice singlet spectral weight: A comprehensive examination by x-ray
  absorption spectroscopy}}, Phys. Rev. B 88 (2013) 134525.
\newblock \href {https://doi.org/10.1103/PhysRevB.88.134525}
  {\path{doi:10.1103/PhysRevB.88.134525}}.
\newline\urlprefix\url{https://link.aps.org/doi/10.1103/PhysRevB.88.134525}

\bibitem{PhysRevLett.115.027002}
N.~B. Brookes, G.~Ghiringhelli, A.-M. Charvet, A.~Fujimori, T.~Kakeshita,
  H.~Eisaki, S.~Uchida, T.~Mizokawa,
  \href{https://link.aps.org/doi/10.1103/PhysRevLett.115.027002}{{Stability of
  the Zhang-Rice Singlet with Doping in Lanthanum Strontium Copper Oxide Across
  the Superconducting Dome and Above}}, Phys. Rev. Lett. 115 (2015) 027002.
\newblock \href {https://doi.org/10.1103/PhysRevLett.115.027002}
  {\path{doi:10.1103/PhysRevLett.115.027002}}.
\newline\urlprefix\url{https://link.aps.org/doi/10.1103/PhysRevLett.115.027002}

\bibitem{PhysRevLett.95.177002}
M.~Taguchi, A.~Chainani, K.~Horiba, Y.~Takata, M.~Yabashi, K.~Tamasaku,
  Y.~Nishino, D.~Miwa, T.~Ishikawa, T.~Takeuchi, K.~Yamamoto, M.~Matsunami,
  S.~Shin, T.~Yokoya, E.~Ikenaga, K.~Kobayashi, T.~Mochiku, K.~Hirata, J.~Hori,
  K.~Ishii, F.~Nakamura, T.~Suzuki,
  \href{https://link.aps.org/doi/10.1103/PhysRevLett.95.177002}{{Evidence for
  Suppressed Screening on the Surface of High Temperature
  ${\mathrm{La}}_{2\ensuremath{-}x}{\mathrm{Sr}}_{x}{\mathrm{CuO}}_{4}$ and
  ${\mathrm{Nd}}_{2\ensuremath{-}x}{\mathrm{Ce}}_{x}{\mathrm{CuO}}_{4}$
  Superconductors}}, Phys. Rev. Lett. 95 (2005) 177002.
\newblock \href {https://doi.org/10.1103/PhysRevLett.95.177002}
  {\path{doi:10.1103/PhysRevLett.95.177002}}.
\newline\urlprefix\url{https://link.aps.org/doi/10.1103/PhysRevLett.95.177002}

\bibitem{Weber2012}
C.~Weber, C.~Yee, K.~Haule, G.~Kotliar,
  \href{http://stacks.iop.org/0295-5075/100/i=3/a=37001?key=crossref.86f7be16c6aa6e8b19bb54661aca90d9}{{Scaling
  of the transition temperature of hole-doped cuprate superconductors with the
  charge-transfer energy}}, EPL (Europhysics Letters) 100~(3) (2012) 37001.
\newblock \href {http://arxiv.org/abs/1108.3028} {\path{arXiv:1108.3028}},
  \href {https://doi.org/10.1209/0295-5075/100/37001}
  {\path{doi:10.1209/0295-5075/100/37001}}.
\newline\urlprefix\url{http://stacks.iop.org/0295-5075/100/i=3/a=37001?key=crossref.86f7be16c6aa6e8b19bb54661aca90d9}

\bibitem{Weber2017102}
C.~Weber,
  \href{http://www.sciencedirect.com/science/article/pii/S2095927316305916}{What
  controls the critical temperature of high temperature copper oxide
  superconductors: insights from scanneling tunnelling microscopy}, Science
  Bulletin 62~(2) (2017) 102.
\newblock \href {https://doi.org/https://doi.org/10.1016/j.scib.2016.12.007}
  {\path{doi:https://doi.org/10.1016/j.scib.2016.12.007}}.
\newline\urlprefix\url{http://www.sciencedirect.com/science/article/pii/S2095927316305916}

\bibitem{Badoux2015}
S.~Badoux, W.~Tabis, F.~Lalibert{\'{e}}, G.~Grissonnanche, B.~Vignolle,
  D.~Vignolles, J.~B{\'{e}}ard, D.~A. Bonn, W.~N. Hardy, R.~Liang,
  N.~Doiron-Leyraud, L.~Taillefer, C.~Proust,
  \href{http://arxiv.org/abs/1511.08162}{{Change of carrier density at the
  pseudogap critical point of a cuprate superconductor}}, Nature 531~(7593)
  (2015) 210.
\newblock \href {http://arxiv.org/abs/1511.08162} {\path{arXiv:1511.08162}},
  \href {https://doi.org/10.1038/nature16983} {\path{doi:10.1038/nature16983}}.
\newline\urlprefix\url{http://arxiv.org/abs/1511.08162}

\bibitem{PhysRevB.95.224517}
C.~Collignon, S.~Badoux, S.~A.~A. Afshar, B.~Michon, F.~Lalibert\'e,
  O.~Cyr-Choini\`ere, J.-S. Zhou, S.~Licciardello, S.~Wiedmann,
  N.~Doiron-Leyraud, L.~Taillefer,
  \href{https://link.aps.org/doi/10.1103/PhysRevB.95.224517}{{Fermi-surface
  transformation across the pseudogap critical point of the cuprate
  superconductor
  ${\mathrm{La}}_{1.6\ensuremath{-}x}{\mathrm{Nd}}_{0.4}{\mathrm{Sr}}_{x}{\mathrm{CuO}}_{4}$}},
  Phys. Rev. B 95 (2017) 224517.
\newblock \href {https://doi.org/10.1103/PhysRevB.95.224517}
  {\path{doi:10.1103/PhysRevB.95.224517}}.
\newline\urlprefix\url{https://link.aps.org/doi/10.1103/PhysRevB.95.224517}

\bibitem{Doiron-Leyraud2017}
N.~Doiron-Leyraud, O.~Cyr-Choini{\`{e}}re, S.~Badoux, A.~Ataei, C.~Collignon,
  A.~Gourgout, S.~Dufour-Beaus{\'{e}}jour, F.~F. Tafti, F.~Lalibert{\'{e}},
  M.-E. Boulanger, M.~Matusiak, D.~Graf, M.~Kim, J.-S. Zhou, N.~Momono,
  T.~Kurosawa, H.~Takagi, L.~Taillefer,
  \href{http://www.nature.com/articles/s41467-017-02122-x}{{Pseudogap phase of
  cuprate superconductors confined by Fermi surface topology}}, Nature
  Communications 8~(1) (2017) 2044.
\newblock \href {http://arxiv.org/abs/1712.05113} {\path{arXiv:1712.05113}},
  \href {https://doi.org/10.1038/s41467-017-02122-x}
  {\path{doi:10.1038/s41467-017-02122-x}}.
\newline\urlprefix\url{http://www.nature.com/articles/s41467-017-02122-x}

\bibitem{Ruan2016}
W.~Ruan, C.~Hu, J.~Zhao, P.~Cai, Y.~Peng, C.~Ye, R.~Yu, X.~Li, Z.~Hao, C.~Jin,
  X.~Zhou, Z.-Y. Weng, Y.~Wang,
  \href{http://dx.doi.org/10.1007/s11434-016-1204-x}{Relationship between the
  parent charge transfer gap and maximum transition temperature in cuprates},
  Science Bulletin 61~(23) (2016) 1826.
\newblock \href {https://doi.org/10.1007/s11434-016-1204-x}
  {\path{doi:10.1007/s11434-016-1204-x}}.
\newline\urlprefix\url{http://dx.doi.org/10.1007/s11434-016-1204-x}

\bibitem{PhysRevB.96.245112}
S.-L. Yang, J.~A. Sobota, Y.~He, Y.~Wang, D.~Leuenberger, H.~Soifer,
  M.~Hashimoto, D.~H. Lu, H.~Eisaki, B.~Moritz, T.~P. Devereaux, P.~S.
  Kirchmann, Z.-X. Shen,
  \href{https://link.aps.org/doi/10.1103/PhysRevB.96.245112}{{Revealing the
  Coulomb interaction strength in a cuprate superconductor}}, Phys. Rev. B 96
  (2017) 245112.
\newblock \href {https://doi.org/10.1103/PhysRevB.96.245112}
  {\path{doi:10.1103/PhysRevB.96.245112}}.
\newline\urlprefix\url{https://link.aps.org/doi/10.1103/PhysRevB.96.245112}

\bibitem{PhysRevB.52.6854}
Y.~Cao, Q.~Xiong, Y.~Y. Xue, C.~W. Chu,
  \href{https://link.aps.org/doi/10.1103/PhysRevB.52.6854}{{Pressure effect on
  the ${\mathit{T}}_{\mathit{c}}$ of
  ${\mathrm{HgBa}}_{2}$${\mathrm{CuO}}_{4+\mathrm{\ensuremath{\delta}}}$ with
  0.07\ensuremath{\le}\ensuremath{\delta}\ensuremath{\le}0.39}}, Phys. Rev. B
  52 (1995) 6854.
\newblock \href {https://doi.org/10.1103/PhysRevB.52.6854}
  {\path{doi:10.1103/PhysRevB.52.6854}}.
\newline\urlprefix\url{https://link.aps.org/doi/10.1103/PhysRevB.52.6854}

\bibitem{Chu1995}
C.~W. Chu, Y.~Cao, Q.~Xiong, Y.~Y. Xue,
  \href{http://dx.doi.org/10.1007/BF00722813}{{$\delta$-Influence on the
  pressure-effect on Tc of $HgBa_2CuO_{4+\delta}$ and the inverse parabolic
  Tc-relation}}, Journal of Superconductivity 8~(4) (1995) 393.
\newblock \href {https://doi.org/10.1007/BF00722813}
  {\path{doi:10.1007/BF00722813}}.
\newline\urlprefix\url{http://dx.doi.org/10.1007/BF00722813}

\bibitem{PhysRevLett.105.167002}
F.~Hardy, N.~J. Hillier, C.~Meingast, D.~Colson, Y.~Li, N.~Bari\ifmmode
  \check{s}\else \v{s}\fi{}i\ifmmode~\acute{c}\else \'{c}\fi{}, G.~Yu, X.~Zhao,
  M.~Greven, J.~S. Schilling,
  \href{http://link.aps.org/doi/10.1103/PhysRevLett.105.167002}{{Enhancement of
  the Critical Temperature of
  ${\mathrm{HgBa}}_{2}{\mathrm{CuO}}_{4+\ensuremath{\delta}}$ by Applying
  Uniaxial and Hydrostatic Pressure: Implications for a Universal Trend in
  Cuprate Superconductors}}, Phys. Rev. Lett. 105 (2010) 167002.
\newblock \href {https://doi.org/10.1103/PhysRevLett.105.167002}
  {\path{doi:10.1103/PhysRevLett.105.167002}}.
\newline\urlprefix\url{http://link.aps.org/doi/10.1103/PhysRevLett.105.167002}

\bibitem{Yamamoto2015}
A.~Yamamoto, N.~Takeshita, C.~Terakura, Y.~Tokura,
  \href{http://www.nature.com/ncomms/2015/151201/ncomms9990/full/ncomms9990.html{\#}ref4{\%}5Cnhttp://www.nature.com/doifinder/10.1038/ncomms9990}{{High
  pressure effects revisited for the cuprate superconductor family with highest
  critical temperature}}, Nature Communications 6 (2015) 8990.
\newblock \href {https://doi.org/10.1038/ncomms9990}
  {\path{doi:10.1038/ncomms9990}}.
\newline\urlprefix\url{http://www.nature.com/ncomms/2015/151201/ncomms9990/full/ncomms9990.html{\#}ref4{\%}5Cnhttp://www.nature.com/doifinder/10.1038/ncomms9990}

\bibitem{PhysRevLett.60.732}
K.~B. Lyons, P.~A. Fleury, L.~F. Schneemeyer, J.~V. Waszczak,
  \href{https://link.aps.org/doi/10.1103/PhysRevLett.60.732}{{Spin fluctuations
  and superconductivity in
  ${\mathrm{Ba}}_{2}$${\mathrm{YCu}}_{3}$${\mathrm{O}}_{6+\mathrm{\ensuremath{\delta}}}$}},
  Phys. Rev. Lett. 60 (1988) 732.
\newblock \href {https://doi.org/10.1103/PhysRevLett.60.732}
  {\path{doi:10.1103/PhysRevLett.60.732}}.
\newline\urlprefix\url{https://link.aps.org/doi/10.1103/PhysRevLett.60.732}

\bibitem{PhysRevB.53.R11930}
G.~Blumberg, P.~Abbamonte, M.~V. Klein, W.~C. Lee, D.~M. Ginsberg, L.~L.
  Miller, A.~Zibold,
  \href{https://link.aps.org/doi/10.1103/PhysRevB.53.R11930}{Resonant
  two-magnon raman scattering in cuprate antiferromagnetic insulators}, Phys.
  Rev. B 53 (1996) R11930.
\newblock \href {https://doi.org/10.1103/PhysRevB.53.R11930}
  {\path{doi:10.1103/PhysRevB.53.R11930}}.
\newline\urlprefix\url{https://link.aps.org/doi/10.1103/PhysRevB.53.R11930}

\bibitem{PhysRevLett.79.4906}
P.~Bourges, H.~Casalta, A.~S. Ivanov, D.~Petitgrand,
  \href{https://link.aps.org/doi/10.1103/PhysRevLett.79.4906}{Superexchange
  coupling and spin susceptibility spectral weight in undoped monolayer
  cuprates}, Phys. Rev. Lett. 79 (1997) 4906.
\newblock \href {https://doi.org/10.1103/PhysRevLett.79.4906}
  {\path{doi:10.1103/PhysRevLett.79.4906}}.
\newline\urlprefix\url{https://link.aps.org/doi/10.1103/PhysRevLett.79.4906}

\bibitem{PhysRevLett.67.3622}
S.~M. Hayden, G.~Aeppli, R.~Osborn, A.~D. Taylor, T.~G. Perring, S.-W. Cheong,
  Z.~Fisk,
  \href{https://link.aps.org/doi/10.1103/PhysRevLett.67.3622}{{High-energy spin
  waves in ${\mathrm{La}}_{2}$${\mathrm{CuO}}_{4}$}}, Phys. Rev. Lett. 67
  (1991) 3622.
\newblock \href {https://doi.org/10.1103/PhysRevLett.67.3622}
  {\path{doi:10.1103/PhysRevLett.67.3622}}.
\newline\urlprefix\url{https://link.aps.org/doi/10.1103/PhysRevLett.67.3622}

\bibitem{PhysRevLett.84.4188}
S.~Daul, D.~J. Scalapino, S.~R. White,
  \href{http://link.aps.org/doi/10.1103/PhysRevLett.84.4188}{Pairing
  correlations on $\mathit{t}\ensuremath{-}\mathit{U}\ensuremath{-}\mathit{J}$
  ladders}, Phys. Rev. Lett. 84 (2000) 4188.
\newblock \href {https://doi.org/10.1103/PhysRevLett.84.4188}
  {\path{doi:10.1103/PhysRevLett.84.4188}}.
\newline\urlprefix\url{http://link.aps.org/doi/10.1103/PhysRevLett.84.4188}

\bibitem{PhysRevB.63.100506}
S.~Basu, R.~J. Gooding, P.~W. Leung,
  \href{https://link.aps.org/doi/10.1103/PhysRevB.63.100506}{Enhanced
  bound-state formation in two dimensions via stripelike hopping anisotropies},
  Phys. Rev. B 63 (2001) 100506.
\newblock \href {https://doi.org/10.1103/PhysRevB.63.100506}
  {\path{doi:10.1103/PhysRevB.63.100506}}.
\newline\urlprefix\url{https://link.aps.org/doi/10.1103/PhysRevB.63.100506}

\bibitem{Laughlin2002}
R.~B. Laughlin, {Gossamer superconductivity}, arXiv:0209269 (2002).

\bibitem{PhysRevLett.90.207002}
F.~C. Zhang,
  \href{http://link.aps.org/doi/10.1103/PhysRevLett.90.207002}{{Gossamer
  Superconductor, Mott Insulator, and Resonating Valence Bond State in
  Correlated Electron Systems}}, Phys. Rev. Lett. 90 (2003) 207002.
\newblock \href {https://doi.org/10.1103/PhysRevLett.90.207002}
  {\path{doi:10.1103/PhysRevLett.90.207002}}.
\newline\urlprefix\url{http://link.aps.org/doi/10.1103/PhysRevLett.90.207002}

\bibitem{Coleman2003}
P.~Coleman, {Superconductivity: lifting the gossamer veil.}, Nature 424 (2003)
  625.
\newblock \href {https://doi.org/10.1038/424625a} {\path{doi:10.1038/424625a}}.

\bibitem{PhysRevB.79.014524}
T.~Xiang, H.~G. Luo, D.~H. Lu, K.~M. Shen, Z.~X. Shen,
  \href{http://link.aps.org/doi/10.1103/PhysRevB.79.014524}{Intrinsic electron
  and hole bands in electron-doped cuprate superconductors}, Phys. Rev. B 79
  (2009) 014524.
\newblock \href {https://doi.org/10.1103/PhysRevB.79.014524}
  {\path{doi:10.1103/PhysRevB.79.014524}}.
\newline\urlprefix\url{http://link.aps.org/doi/10.1103/PhysRevB.79.014524}

\bibitem{PhysRevLett.90.187004}
E.~Plekhanov, S.~Sorella, M.~Fabrizio,
  \href{https://link.aps.org/doi/10.1103/PhysRevLett.90.187004}{Increasing
  $d$-wave superconductivity by on-site repulsion}, Phys. Rev. Lett. 90 (2003)
  187004.
\newblock \href {https://doi.org/10.1103/PhysRevLett.90.187004}
  {\path{doi:10.1103/PhysRevLett.90.187004}}.
\newline\urlprefix\url{https://link.aps.org/doi/10.1103/PhysRevLett.90.187004}

\bibitem{PhysRevB.72.045130}
Y.~Wen, Y.~Yu,
  \href{https://link.aps.org/doi/10.1103/PhysRevB.72.045130}{{One-band Hubbard
  model with hopping asymmetry and the effective theory at finite $U$: Phase
  diagram and metal-insulator transition}}, Phys. Rev. B 72 (2005) 045130.
\newblock \href {https://doi.org/10.1103/PhysRevB.72.045130}
  {\path{doi:10.1103/PhysRevB.72.045130}}.
\newline\urlprefix\url{https://link.aps.org/doi/10.1103/PhysRevB.72.045130}

\bibitem{PhysRevB.79.144526}
S.~Guertler, Q.-H. Wang, F.-C. Zhang,
  \href{https://link.aps.org/doi/10.1103/PhysRevB.79.144526}{Variational monte
  carlo studies of gossamer superconductivity}, Phys. Rev. B 79 (2009) 144526.
\newblock \href {https://doi.org/10.1103/PhysRevB.79.144526}
  {\path{doi:10.1103/PhysRevB.79.144526}}.
\newline\urlprefix\url{https://link.aps.org/doi/10.1103/PhysRevB.79.144526}

\bibitem{PhysRevB.90.115137}
T.~Misawa, M.~Imada,
  \href{http://link.aps.org/doi/10.1103/PhysRevB.90.115137}{Origin of
  high-${T}_{c}$ superconductivity in doped hubbard models and their
  extensions: Roles of uniform charge fluctuations}, Phys. Rev. B 90 (2014)
  115137.
\newblock \href {https://doi.org/10.1103/PhysRevB.90.115137}
  {\path{doi:10.1103/PhysRevB.90.115137}}.
\newline\urlprefix\url{http://link.aps.org/doi/10.1103/PhysRevB.90.115137}

\bibitem{1605.03969}
W.-C. Lee, {Crossover From Strong to Weak Pairing States in t-J-U Model Studied
  by A Slave Spin Method}, arXiv:1605.03969 (2016).

\bibitem{0953-8984-28-11-116001}
S.~Acharya, A.~Medhi, N.~S. Vidhyadhiraja, A.~Taraphder,
  \href{http://stacks.iop.org/0953-8984/28/i=11/a=116001}{Feasibility of a
  metamagnetic transition in correlated systems}, Journal of Physics: Condensed
  Matter 28~(11) (2016) 116001.
\newline\urlprefix\url{http://stacks.iop.org/0953-8984/28/i=11/a=116001}

\bibitem{PhysRevB.71.014508}
J.~Y. Gan, F.~C. Zhang, Z.~B. Su,
  \href{https://link.aps.org/doi/10.1103/PhysRevB.71.014508}{Theory of gossamer
  and resonating valence bond superconductivity}, Phys. Rev. B 71 (2005)
  014508.
\newblock \href {https://doi.org/10.1103/PhysRevB.71.014508}
  {\path{doi:10.1103/PhysRevB.71.014508}}.
\newline\urlprefix\url{https://link.aps.org/doi/10.1103/PhysRevB.71.014508}

\bibitem{PhysRevB.71.104505}
F.~Yuan, Q.~Yuan, C.~S. Ting,
  \href{http://link.aps.org/doi/10.1103/PhysRevB.71.104505}{{Gossamer
  superconductivity and antiferromagnetism in the
  $t\text{\penalty1000-\hskip0pt}J\text{\penalty1000-\hskip0pt}U$ model}},
  Phys. Rev. B 71 (2005) 104505.
\newblock \href {https://doi.org/10.1103/PhysRevB.71.104505}
  {\path{doi:10.1103/PhysRevB.71.104505}}.
\newline\urlprefix\url{http://link.aps.org/doi/10.1103/PhysRevB.71.104505}

\bibitem{PhysRevB.88.094502}
M.~Abram, J.~Kaczmarczyk, J.~J\ifmmode~\mbox{\k{e}}\fi{}drak, J.~Spa\l{}ek,
  \href{http://link.aps.org/doi/10.1103/PhysRevB.88.094502}{{d-wave
  superconductivity and its coexistence with antiferromagnetism in the
  t–J–U model: Statistically consistent Gutzwiller approach}}, Phys. Rev. B
  88 (2013) 094502.
\newblock \href {https://doi.org/10.1103/PhysRevB.88.094502}
  {\path{doi:10.1103/PhysRevB.88.094502}}.
\newline\urlprefix\url{http://link.aps.org/doi/10.1103/PhysRevB.88.094502}

\bibitem{Liu2014123}
B.~Liu, X.~Yan, F.~Yuan,
  \href{http://www.sciencedirect.com/science/article/pii/S0038109813004900}{Quasiparticle
  resonance states induced by a nonmagnetic impurity in gossamer
  superconductors}, Solid State Communications 177 (2014) 123.
\newblock \href {https://doi.org/https://doi.org/10.1016/j.ssc.2013.10.016}
  {\path{doi:https://doi.org/10.1016/j.ssc.2013.10.016}}.
\newline\urlprefix\url{http://www.sciencedirect.com/science/article/pii/S0038109813004900}

\bibitem{0253-6102-57-4-29}
L.~Fen-Fen, Z.~Yong, Y.~Feng, X.~Lin-Hua,
  \href{http://stacks.iop.org/0253-6102/57/i=4/a=29}{Effects of the next
  nearest neighbor hopping on superconductivity and antiferromagnetism of
  gossamer superconductivity}, Communications in Theoretical Physics 57~(4)
  (2012) 727.
\newline\urlprefix\url{http://stacks.iop.org/0253-6102/57/i=4/a=29}

\bibitem{0953-8984-23-49-495602}
K.-K. Voo, \href{http://stacks.iop.org/0953-8984/23/i=49/a=495602}{{Order and
  excitation in partially Gutzwiller projected $t - t^{\prime}-
  t^{\prime\prime}- J - U $models}}, Journal of Physics: Condensed Matter
  23~(49) (2011) 495602.
\newline\urlprefix\url{http://stacks.iop.org/0953-8984/23/i=49/a=495602}

\bibitem{Abram_2017}
M.~Abram, M.~Zegrodnik, J.~Spa{\l}ek,
  \href{https://doi.org/10.1088%2F1361-648x%2Faa7a21}{{Antiferromagnetism,
  charge density wave, and d-wave superconductivity in the extended t-J-U
  model: role of intersite Coulomb interaction and a critical overview of
  renormalized mean field theory}}, Journal of Physics: Condensed Matter
  29~(36) (2017) 365602.
\newblock \href {https://doi.org/10.1088/1361-648x/aa7a21}
  {\path{doi:10.1088/1361-648x/aa7a21}}.
\newline\urlprefix\url{https://doi.org/10.1088%2F1361-648x%2Faa7a21}

\bibitem{PhysRevB.96.054511}
M.~Zegrodnik, J.~Spa\l{}ek,
  \href{https://link.aps.org/doi/10.1103/PhysRevB.96.054511}{{Universal
  properties of high-temperature superconductors from real-space pairing: Role
  of correlated hopping and intersite Coulomb interaction within the $t-J-U$
  model}}, Phys. Rev. B 96 (2017) 054511.
\newblock \href {https://doi.org/10.1103/PhysRevB.96.054511}
  {\path{doi:10.1103/PhysRevB.96.054511}}.
\newline\urlprefix\url{https://link.aps.org/doi/10.1103/PhysRevB.96.054511}

\bibitem{PhysRevB.95.024507}
M.~Zegrodnik, J.~Spa\l{}ek,
  \href{https://link.aps.org/doi/10.1103/PhysRevB.95.024507}{{Effect of
  interlayer processes on the superconducting state within the
  $t\ensuremath{-}J\ensuremath{-}U$ model: Full Gutzwiller wave-function
  solution and relation to experiment}}, Phys. Rev. B 95 (2017) 024507.
\newblock \href {https://doi.org/10.1103/PhysRevB.95.024507}
  {\path{doi:10.1103/PhysRevB.95.024507}}.
\newline\urlprefix\url{https://link.aps.org/doi/10.1103/PhysRevB.95.024507}

\bibitem{PhysRevB.95.024506}
J.~Spa\l{}ek, M.~Zegrodnik, J.~Kaczmarczyk,
  \href{https://link.aps.org/doi/10.1103/PhysRevB.95.024506}{{Universal
  properties of high-temperature superconductors from real-space pairing:
  $t\ensuremath{-}J\ensuremath{-}U$ model and its quantitative comparison with
  experiment}}, Phys. Rev. B 95 (2017) 024506.
\newblock \href {https://doi.org/10.1103/PhysRevB.95.024506}
  {\path{doi:10.1103/PhysRevB.95.024506}}.
\newline\urlprefix\url{https://link.aps.org/doi/10.1103/PhysRevB.95.024506}

\bibitem{PhysRevB.96.205120}
A.~Nocera, N.~D. Patel, E.~Dagotto, G.~Alvarez,
  \href{https://link.aps.org/doi/10.1103/PhysRevB.96.205120}{Signatures of
  pairing in the magnetic excitation spectrum of strongly correlated two-leg
  ladders}, Phys. Rev. B 96 (2017) 205120.
\newblock \href {https://doi.org/10.1103/PhysRevB.96.205120}
  {\path{doi:10.1103/PhysRevB.96.205120}}.
\newline\urlprefix\url{https://link.aps.org/doi/10.1103/PhysRevB.96.205120}

\bibitem{YHL2016}
Y.-H. Liu, \href{https://hdl.handle.net/11296/wz6ru7}{{Charge transfer model :
  What is the effect of charge fluctuation in t-J model}}, National Tsing Hua
  University~(Master thesis) (2016).
\newline\urlprefix\url{https://hdl.handle.net/11296/wz6ru7}

\bibitem{refId0}
{Bouchaud, J.P.}, {Georges, A.}, {Lhuillier, C.},
  \href{https://doi.org/10.1051/jphys:01988004904055300}{Pair wave functions
  for strongly correlated fermions and their determinantal representation}, J.
  Phys. France 49~(4) (1988) 553.
\newblock \href {https://doi.org/10.1051/jphys:01988004904055300}
  {\path{doi:10.1051/jphys:01988004904055300}}.
\newline\urlprefix\url{https://doi.org/10.1051/jphys:01988004904055300}

\bibitem{doi:10.1143/JPSJ.65.3615}
H.~Yokoyama, M.~Ogata, \href{https://doi.org/10.1143/JPSJ.65.3615}{{Phase
  Diagram and Pairing Symmetry of the Two-Dimensional t- J Model by a Variation
  Theory}}, Journal of the Physical Society of Japan 65~(11) (1996) 3615.
\newblock \href {http://arxiv.org/abs/https://doi.org/10.1143/JPSJ.65.3615}
  {\path{arXiv:https://doi.org/10.1143/JPSJ.65.3615}}, \href
  {https://doi.org/10.1143/JPSJ.65.3615} {\path{doi:10.1143/JPSJ.65.3615}}.
\newline\urlprefix\url{https://doi.org/10.1143/JPSJ.65.3615}

\bibitem{PhysRevB.43.12943}
T.~Giamarchi, C.~Lhuillier,
  \href{https://link.aps.org/doi/10.1103/PhysRevB.43.12943}{{Phase diagrams of
  the two-dimensional Hubbard and t-J models by a variational Monte Carlo
  method}}, Phys. Rev. B 43 (1991) 12943.
\newblock \href {https://doi.org/10.1103/PhysRevB.43.12943}
  {\path{doi:10.1103/PhysRevB.43.12943}}.
\newline\urlprefix\url{https://link.aps.org/doi/10.1103/PhysRevB.43.12943}

\bibitem{Shih2004}
C.~T. Shih, T.~K. Lee, R.~Eder, C.-Y. Mou, Y.~C. Chen,
  \href{http://link.aps.org/doi/10.1103/PhysRevLett.92.227002}{{Enhancement of
  Pairing Correlation by ${t}^{\ensuremath{'}}$ in the Two-Dimensional Extended
  $t-J$ Model}}, Phys. Rev. Lett. 92 (2004) 227002.
\newblock \href {https://doi.org/10.1103/PhysRevLett.92.227002}
  {\path{doi:10.1103/PhysRevLett.92.227002}}.
\newline\urlprefix\url{http://link.aps.org/doi/10.1103/PhysRevLett.92.227002}

\bibitem{PhysRevB.70.220502}
C.~T. Shih, Y.~C. Chen, C.~P. Chou, T.~K. Lee,
  \href{https://link.aps.org/doi/10.1103/PhysRevB.70.220502}{{Absence of the
  coexistence of superconductivity and antiferromagnetism in the hole-doped
  two-dimensional extended $t\text{\ensuremath{-}}J$ model}}, Phys. Rev. B 70
  (2004) 220502.
\newblock \href {https://doi.org/10.1103/PhysRevB.70.220502}
  {\path{doi:10.1103/PhysRevB.70.220502}}.
\newline\urlprefix\url{https://link.aps.org/doi/10.1103/PhysRevB.70.220502}

\bibitem{PhysRevLett.10.159}
M.~C. Gutzwiller,
  \href{http://link.aps.org/doi/10.1103/PhysRevLett.10.159}{Effect of
  correlation on the ferromagnetism of transition metals}, Phys. Rev. Lett. 10
  (1963) 159.
\newblock \href {https://doi.org/10.1103/PhysRevLett.10.159}
  {\path{doi:10.1103/PhysRevLett.10.159}}.
\newline\urlprefix\url{http://link.aps.org/doi/10.1103/PhysRevLett.10.159}

\bibitem{PhysRev.98.1479}
R.~Jastrow, \href{https://link.aps.org/doi/10.1103/PhysRev.98.1479}{Many-body
  problem with strong forces}, Phys. Rev. 98 (1955) 1479--1484.
\newblock \href {https://doi.org/10.1103/PhysRev.98.1479}
  {\path{doi:10.1103/PhysRev.98.1479}}.
\newline\urlprefix\url{https://link.aps.org/doi/10.1103/PhysRev.98.1479}

\bibitem{doi:10.7566/JPSCP.3.015012}
K.~Kobayashi, H.~Yokoyama,
  \href{https://journals.jps.jp/doi/abs/10.7566/JPSCP.3.015012}{{Coexistence
  and Mutual Exclusion of Superconductivity and Antiferromagnetism, and Phase
  Separation in Hubbard Model}}, Proceedings of the International Conference on
  Strongly Correlated Electron Systems (2013).
\newblock \href
  {http://arxiv.org/abs/https://journals.jps.jp/doi/pdf/10.7566/JPSCP.3.015012}
  {\path{arXiv:https://journals.jps.jp/doi/pdf/10.7566/JPSCP.3.015012}}, \href
  {https://doi.org/10.7566/JPSCP.3.015012} {\path{doi:10.7566/JPSCP.3.015012}}.
\newline\urlprefix\url{https://journals.jps.jp/doi/abs/10.7566/JPSCP.3.015012}

\bibitem{doi:10.7566/JPSJ.82.014707}
H.~Yokoyama, M.~Ogata, Y.~Tanaka, K.~Kobayashi, H.~Tsuchiura,
  \href{http://dx.doi.org/10.7566/JPSJ.82.014707}{{Crossover between BCS
  Superconductor and Doped Mott Insulator of d-Wave Pairing State in
  Two-Dimensional Hubbard Model}}, Journal of the Physical Society of Japan
  82~(1) (2013) 014707.
\newblock \href {http://arxiv.org/abs/http://dx.doi.org/10.7566/JPSJ.82.014707}
  {\path{arXiv:http://dx.doi.org/10.7566/JPSJ.82.014707}}, \href
  {https://doi.org/10.7566/JPSJ.82.014707} {\path{doi:10.7566/JPSJ.82.014707}}.
\newline\urlprefix\url{http://dx.doi.org/10.7566/JPSJ.82.014707}

\bibitem{PhysRevB.95.035133}
H.-K. Wu, T.-K. Lee,
  \href{http://link.aps.org/doi/10.1103/PhysRevB.95.035133}{{Spectral evolution
  with doping of an antiferromagnetic Mott state}}, Phys. Rev. B 95 (2017)
  035133.
\newblock \href {https://doi.org/10.1103/PhysRevB.95.035133}
  {\path{doi:10.1103/PhysRevB.95.035133}}.
\newline\urlprefix\url{http://link.aps.org/doi/10.1103/PhysRevB.95.035133}

\bibitem{PhysRevB.71.241103}
S.~Sorella, \href{http://link.aps.org/doi/10.1103/PhysRevB.71.241103}{{Wave
  function optimization in the variational Monte Carlo method}}, Phys. Rev. B
  71 (2005) 241103.
\newblock \href {https://doi.org/10.1103/PhysRevB.71.241103}
  {\path{doi:10.1103/PhysRevB.71.241103}}.
\newline\urlprefix\url{http://link.aps.org/doi/10.1103/PhysRevB.71.241103}

\bibitem{Weber2010a}
C.~Weber, K.~Haule, G.~Kotliar,
  \href{http://www.nature.com/nphys/journal/v6/n8/full/nphys1706.html{\#}/ref22{\%}5Cnhttp://www.nature.com/nphys/journal/v6/n8/pdf/nphys1706.pdf}{{Strength
  of correlations in electron- and hole-doped cuprates}}, Nature Physics 6~(8)
  (2010) 574--578.
\newblock \href {https://doi.org/10.1038/nphys1706}
  {\path{doi:10.1038/nphys1706}}.
\newline\urlprefix\url{http://www.nature.com/nphys/journal/v6/n8/full/nphys1706.html{\#}/ref22{\%}5Cnhttp://www.nature.com/nphys/journal/v6/n8/pdf/nphys1706.pdf}

\bibitem{Weber2010}
C.~Weber, K.~Haule, G.~Kotliar, {Apical oxygens and correlation strength in
  electron- and hole-doped copper oxides}, Physical Review B - Condensed Matter
  and Materials Physics 82~(12) (2010) 1--24.
\newblock \href {http://arxiv.org/abs/1005.3100} {\path{arXiv:1005.3100}},
  \href {https://doi.org/10.1103/PhysRevB.82.125107}
  {\path{doi:10.1103/PhysRevB.82.125107}}.

\bibitem{Cai2016}
P.~Cai, W.~Ruan, Y.~Peng, C.~Ye, X.~Li, Z.~Hao, X.~Zhou, D.-H. Lee, Y.~Wang,
  \href{http://dx.doi.org/10.1038/nphys3840}{Visualizing the evolution from the
  mott insulator to a charge-ordered insulator in lightly doped cuprates}, Nat
  Phys 12~(11) (2016) 1047--1051, letter.
\newline\urlprefix\url{http://dx.doi.org/10.1038/nphys3840}

\end{thebibliography}

\end{document}